\documentclass[preprint]{aastex631}

\usepackage{amsmath}
\usepackage{ulem}
\usepackage{color}

\usepackage{natbib}
\submitjournal{ApJ}

\shorttitle{Quantitative Study of CME Magnetic Flux Rope Topology}
\shortauthors{He et al.}

\graphicspath{{./}{figures/}}

\begin{document}

% \title{Quantitative study of CME Magnetic Flux Rope topology in Active Region 11719 and 12158}
\title{Quantitative Characterization of Magnetic Flux Rope Properties for Two Solar Eruption Events}% ??
% \title{Quantitative investigation of two CME Magnetic Flux Ropes on the Sun and their in situ counterparts}

\correspondingauthor{Qiang Hu}
\email{qh0001@uah.edu}

\author[0000-0001-8749-1022]{Wen He}
\affiliation{Department of Space Science,\\
%Center for Space Plasma and Aeronomic Research (CSPAR), \\
The University of Alabama in Huntsville, \\
Huntsville, AL 35805, USA}

\author[0000-0002-7570-2301]{Qiang Hu}
\affiliation{Department of Space Science,\\
and Center for Space Plasma and Aeronomic Research (CSPAR), \\
The University of Alabama in Huntsville, \\
Huntsville, AL 35805, USA}

\author[0000-0002-7018-6862]{Chaowei Jiang}
\affiliation{Institute of Space Science and Applied Technology, \\
Harbin Institute of Technology, \\
Shenzhen, China }

\author{Jiong Qiu}
\affiliation{Physics Department, Montana State University, Bozeman, MT 59717, USA}

\author[0000-0003-0819-464X]{Avijeet Prasad}
\affiliation{Institute of Theoretical Astrophysics, \\
University of Oslo, Postboks 1029 Blindern, 0315 Oslo, Norway}
\affiliation{Rosseland Centre for Solar Physics, \\
University of Oslo, Postboks 1029 Blindern, 0315 Oslo, Norway}

\begin{abstract}

In order to bridge the gap between heliospheric and solar observations of coronal mass ejections (CMEs), one of the key steps is to improve the understanding of their corresponding magnetic structures like the magnetic flux ropes (MFRs). But it remains a challenge to confirm the existence of a coherent MFR before or upon the CME eruption on the Sun and to quantitatively characterize the CME-MFR due to the lack of direct magnetic field measurement in the corona. In this study, we investigate the MFR structures, originating from two active regions (ARs), AR 11719 and AR 12158, and estimate their magnetic properties quantitatively. We perform the nonlinear force-free field extrapolations with preprocessed photospheric vector magnetograms. In addition, remote-sensing observations are employed to find indirect evidence of MFRs on the Sun and to analyze the time evolution of magnetic reconnection flux associated with the flare ribbons during the eruption. A coherent ``pre-existing'' MFR structure prior to the flare eruption is  identified quantitatively for one event from the combined analysis of the extrapolation and observation. Then the characteristics of MFRs for two events on the Sun before and during the eruption, forming the CME-MFR, including the axial magnetic flux, field-line twist, and reconnection flux, are estimated and compared with the corresponding in situ modeling results. We find that the magnetic reconnection associated with the accompanying flares for both events injects significant amount of flux into the erupted CME-MFRs.

\end{abstract}



\section{Introduction} \label{0}

Coronal mass ejections (CMEs) are spectacular eruptions of plasma often accompanied with rapid release of magnetic energy from solar atmosphere. When CMEs propagate away from the Sun to the interplanetary space, they are often called interplanetary CMEs (ICMEs) which exhibit a distinct set of observational signatures from in situ measurements. They are recognized as drivers of major space weather events which could severely impact human activities in modern society. Erupted flares have a close relationship with CMEs and strong flares are often observed along with CMEs \citep{2011Chen}. In the past few decades, a lot of efforts have been made on the development of the flare-CME model in order to explain the underlying physical mechanism(s). Among them, the standard two-dimensional (2D) flare and CME model, so-called CSHKP model \citep[developed by][]{1964Carmichael, 1966Sturrock, 1974Hirayama, 1976KoppPneuman}, has successfully explained the morphological evolution of eruptive two-ribbon flares. One of the essential components in the model is a CME magnetic flux rope (MFR). Upon its ejection, field lines below the erupting rope reconnect and form an arcade of flare loops observed in the X-ray and extreme ultraviolet (EUV) wavelengths, and two ribbons observed in optical and ultraviolet (UV) wavelengths demarcating the feet of the arcade. In the 2D model, the same amount of magnetic flux encompassed by the flare loops is also turned into the poloidal flux of the erupting MFR, and this amount of flux can be measured from flare ribbons sweeping through photospheric magnetic field \citep{1984Forbes, 2004Qiu, 2007Qiu}. Since MFRs are generally believed to be the core magnetic structure of CMEs and ICMEs~\citep[e.g.,][]{2006Gibson, 2014Hu, 2017Gopalswamy, 2018Gopalswamy, 2020Liurui}, studies of MFR properties related to their formation and evolution remain an important topic to be explored. We note that hereafter we use ``MFR" for a more generic reference to a magnetic flux rope at different stages of evolution, and ``CME-MFR" for the specific reference of a magnetic flux rope at the final stage of a CME eruption (i.e., when the MFR has well formed after the flare reconnection).

Nowadays, the study of evolution and propagation of CMEs/ICMEs between Sun and Earth is greatly advanced with the help of multiple measurements from multi-spacecraft missions. The large-scale magnetic clouds (MCs) in ICMEs which are commonly detected in situ provide  direct evidence for the existence of erupted CME-MFRs that come from the Sun~\citep{1992burlaga,2007Qiu,2014Hu}. In addition, there are also remote-sensing observations available throughout the interplanetary space. For example,  the twin spacecraft, Solar TErrestrial RElations Observatory (STEREO, \citealt{2008Kaiser}),  can trace CMEs from the high corona to the inner heliosphere via coronagraphic observations. The STEREO mission provides two more viewpoints towards the Sun, in addition to the viewpoint from Earth provided by the Solar and Heliospheric Observatory (SOHO), in the past decades. There are also various signatures for MFRs on the Sun from remote-sensing observations, including filaments, coronal cavities, sigmoids, and hot channels \citep{2017Cheng}. These solar phenomena can be unified into one framework as distinct manifestations of MFRs \citep{2020Liurui}. Most of the latest recognized observational features are attributed to the observations from the Solar Dynamics Observatory (SDO, \citealt{2012Pesnell}). Recent development of large ground-based solar telescopes also becomes an indispensable way to reveal the fine-scale structures and dynamics of MFR formation in the low corona \citep[e.g.,][]{2015Wang}.

Compared to various studies based on in situ measurements of MFR structures after the eruption, the origination of CME-MFRs before and during eruptions still remains elusive due to the complex environment in the solar source region and  limited observations. At the present time, there are certain hypotheses on the formation process of MFRs. Some studies indicate that MFRs could exist prior to the eruption. For example, both \citet{2001Fan} and \citet{2004Magara} reported magnetohydrodynamic (MHD) simulation results that a twisted MFR initially formed below the photosphere can partially emerge into the low corona by magnetic buoyancy. While some studies suggest that the presence of pre-eruptive MFRs is not necessary and MFRs could be built up in the corona via magnetic reconnection processes associated with flares \citep{2003Amari, 2001Moore, 1999Antiochos, 2021Jiang2, 2021Jiang}. To understand the physical processes more precisely for the flare-CME events, extensions of the standard 2D flare model have been proposed to account for much broader ranges of quantitative measurements with three-dimensional (3D) features intrinsic to realistic solar eruptions \citep{2007Longcope, 2012Aulanier, 2017Priest, 2019Aulanier}. For example, quasi-3D models have been developed with a non-vanishing magnetic field component along the axis of the MFR and to illustrate a scenario that sequential reconnection along magnetic polarity inversion line (PIL) forms the MFR in the first place \citep{1989vanBallegooijen, 2007Longcope,2015Schmieder}. This scenario has been widely applied to infer and interpret the magnetic reconnection properties based on the observed flare ribbon morphology \citep{2002Qiu, 2004Qiu, 2010Qiu, 2014Hu, 2017Kazachenko, 2020Zhu}. From such analyses, \citet{2004Qiu} illustrated that there is a temporal correlation between the magnetic reconnection rate and the acceleration of the CME (considered as the eruptive MFR) in the low corona. Such a correlation has been further established by \citet{2020Zhu} based on a statistical study of $\sim$ 60 events. In addition, \citet{2007Qiu} and \citet{2014Hu} showed the correlation between the magnetic reconnection flux and the flux contents of the corresponding ICME/MC flux ropes based on modeling results employing in situ spacecraft measurements. These results support the hypothesis, especially concerning CME-MFRs, that the CME-MFR can be formed by magnetic reconnection during the corresponding flare process. And recent simulation results also indicate clearly that the reconnection flux contributes to the axial (toroidal) flux of the CME-MFR in the early stage \citep{2021Jiang2, 2018Satoshi}.

For the quantitative MFR identification in the solar source region, numerical models can be applied to find MFRs in addition to observations. For the topological analysis of a solar MFR, the 3D magnetic field configuration is commonly obtained through coronal magnetic field extrapolation  based on photospheric magnetograms. A number of high resolution extrapolation results employing different numerical schemes has been compared to observations to study the properties of MFRs in the  magnetically dominant environment on the Sun \citep{2008Schrijver, 2008Thalmann, 2009Wheatland, 2009DeRosa, 2012Wiegelmannb, 2012Sun, 2014Jiang2, 2016Guob}. Among many extrapolation studies,  the force-free approximation is commonly adopted for the case of low plasma $\beta$ (the ratio between the plasma pressure and magnetic pressure) over certain heights in the solar atmosphere above the photosphere \citep{2001Gary}. Under this assumption, the non-magnetic forces including the inertial force can be ignored for a static and time stationary equilibrium. Therefore, the Lorentz force should be self-balanced, and it should satisfy the equation, $\mathbf{J} \times \mathbf{B} = 0$, which means that the electric current density \textbf{J} is parallel to the magnetic field \textbf{B}, with $\mathbf{J} = \alpha \mathbf{B}$ ($\alpha$ is the so-called force-free parameter). The simplest case is the potential field when $\alpha \equiv 0$. If $\alpha$ is not zero, there are two situations depending on whether $\alpha$ is constant or varying in space. One is the linear force-free field (LFFF) for $\alpha \equiv const$ and the other is the nonlinear force-free field (NLFFF). Since our interest lies in the magnetic structure in the ARs on a local scale with high spatial resolution, the most common and practical way to reconstruct the coronal magnetic field is the NLFFF extrapolation method. 

There are a variety of numerical methods proposed to reconstruct the NLFFF for an AR from boundary conditions and sometimes pseudo-initial conditions, including the upward integration, Grad-Rubin iteration, MHD relaxation, optimization approach, and so on \citep[see a review by][]{2021Wiegelmann}. The computation speed and accuracy of different numerical methods can vary significantly given the differences in algorithms and their specific realizations in many aspects. We apply a kind of MHD-relaxation method with a conservation-element/solution-element (CESE) solver \citep{2011Jiang}. The so-called CESE-MHD-NLFFF code \citep{2013Jiang} has been tested by different benchmark cases \citep{1990Low, 2007vanBallegooijen, 2008Metcalf}. And it has also been widely applied to the extrapolation of realistic solar magnetic field data \citep{2013Jiang, 2014Jiang2, 2017Duan, 2019Duan}.

The understanding towards the formation and evolution process of the CME-MFRs will ultimately help us  make a definitive and physical connection between the origin of solar MFRs (including the MFR before and after the eruption) and their interplanetary counterparts. Such a connection can be pursued through a quantitative study of MFR's physical characteristics (e.g., magnetic flux, field-line twist, and electric current). Specifically, one critical step is the detailed analysis of available solar observations in order to determine whether an MFR exists or how one can form prior to and during the eruption. Characterization of MFR properties not only plays a major role in understanding the physical mechanisms underlying solar eruption and the subsequent evolution, but also contributes to the improvement of the forecast ability in space weather. 

In this paper, we carry out the coronal magnetic field extrapolation based on the method developed by \citet{2013Jiang} for two events to obtain 3D magnetic field topology of the AR prior to eruption. In addition to the magnetic field extrapolation, we estimate the possible locations of the MFR's footpoints prior to eruption and measure a number of magnetic field parameters (including the magnetic reconnection flux derived from flare ribbons) in the corresponding AR during the eruption process from different observations. Then the results from extrapolation and observation are combined to identify whether there is a coherent pre-existing MFR before the eruption and to interpret how such a structure may evolve during and following the corresponding flare and CME eruption process. The magnetic properties of the CME-MFR will be further analyzed and compared with the in situ ICME/MC modeling results which are obtained separately. 

This article is organized as follows. The two selected events and extrapolation method are introduced in Section \ref{1}. Then we describe the identification methods and analyze results associated with the MFRs on the Sun in Section \ref{2}. The magnetic properties of MFRs are estimated quantitatively and presented in Section \ref{3} based on results from both the solar source region and in situ modeling. The conclusion is given in Section \ref{4}.

\section{Events Selection and Extrapolation Method}\label{1}
\subsection{Events Overview}
For the purpose of performing a quantitative study of the CME-MFR, we search for appropriate event candidates from a list of reconstructed MFRs based on photospheric magnetograms before the eruption by \citet{2019Duan}. They extrapolated the 3D magnetic field in the active regions for 45 major flare eruption events employing the NLFFF extrapolation code by \citet{2013Jiang}. With a set of criteria similar to those by \citet{2018Jing}, all major flares that are above GOES-class M5 and occurred within $45\degr$ of the solar disk center from 2011 January to 2017 December are selected. Moreover, we also examine the associations between flare and CME, and between CME and ICME to ensure that there exists a well-established one-to-one connection among a flare, an associated CME and an ICME based on the work by \citet{2020Zhu}. Two events are selected for this study as shown in Table \ref{twoevent}. Both ARs related to the two events  located near the disk center. The CMEs associated with the corresponding flares in these two events have been both observed by SOHO and STEREO spacecraft close to the peak times of the corresponding flares \citep{2014Vemareddy,2015Cheng,2016Cheng,2017Joshi}. And the associated ICME/MC events were also observed by the WIND and ACE spacecraft at 1 au, which have been reported by \citet{2021Hu} and \citet{2021Kilpua}, respectively. 

In event 1, an M6.5 flare was produced at $\sim$ 6:55 UT on 2013 April 11 (N07E13). Then a halo CME appeared in the field of view of SOHO/LASCO after 07:24 UT, and the same CME was also observed simultaneously by both STEREO A and B spacecraft after $\sim$ 07:39 UT. The corresponding ICME/MC passing Earth was detected about three days later \citep{2021Hu}. Similar examination is conducted for the second event, which started with an X1.6 flare peaking at $\sim$ 17:45 UT on 2014 September 10. There was also a halo CME following the flare based on the observations from SOHO/LASCO and the coronagraph of STEREO B (data from STEREO A was unavailable during event 2). After two days, the WIND spacecraft encountered the subsequent ICME/MC structure at Earth \citep{2021Kilpua}. Therefore, the connections of CME-MFRs from the Sun to Earth are well established for these two events. We will mainly present the quantitative study of the CME-MFR before the eruption hereafter.

\subsection{CESE-NLFFF-MHD Extrapolation Method}

The CESE-MHD-NLFFF code solves the MHD momentum equation and the magnetic induction equation iteratively until a stationary magnetic field solution is reached, similar to a magnetofrictional approach. As a special case of the MHD relaxation method, the magnetofrictional method includes an artificial dissipative term $\textbf{D(v)}$ to balance the momentum equation with flow velocity \textbf{v}. Specifically, in the CESE-MHD-NLFFF code, $\textbf{D(v)}$ is written in a frictional form $\nu \rho \textbf{v}$ \citep[see below, and][]{2012Jiang} along with some modifications in order to utilize the existing CESE-MHD solver. The modified momentum equation and the induction equation are written as \citep{2012Jiang, 2013Jiang},
\begin{equation}\label{eq1}
  \frac{\partial (\rho \textbf{v})}{\partial t} =(\nabla \times \textbf{B})\times \textbf{B} -\nu \rho \textbf{v}, \rho =|\textbf{B}|^{2} +\rho_{0}.
\end{equation}
\begin{equation}\label{eq2}
  \frac{\partial \textbf{B}}{\partial t} = \nabla \times(\textbf{v} \times \textbf{B}) 
  -\textbf{v} \nabla \cdot \textbf{B} + \nabla(\mu \nabla \cdot \textbf{B}).
\end{equation}
Here the (pseudo) mass density $\rho$ also contains the constant $\mu_{0}$ for simplicity and is assumed to be largely proportional to $|\textbf{B}|^{2}$ in order to roughly equalize the speed of the evolution of the entire field with nearly uniform Alfv\'en speed ($v_{A} =\frac{|\textbf{B}|}{\sqrt{\rho}} \approx const$). To enhance the ability of handling noisy data in realistic solar magnetograms, a small value $\rho_{0}$ is added, e.g., $\rho_{0}$ = 0.1 (in the same unit as $|\textbf{B}|^{2}$), to the original pseudo density $\rho$. Two extra terms are added to control the divergence of the magnetic field in the induction equation. More details can be found in \citet{2012Jiang, 2013Jiang}.

\subsection{Data Preprocessing and Grid Initialization}
Passing across the inhomogeneous plasma environment, the plasma $\beta$ could vary from $\beta >1$ in the photosphere to $\beta \ll 1$ in the low and middle corona, and to $\beta >1 $ again in the upper corona \citep*{1989Gary, 2001Gary}. So the force-free condition may not be always satisfied especially at the photosphere \citep{1995Metcalf}. One way to get a more consistent boundary condition for NLFFF extrapolation is to modify the raw photospheric magnetogram to mimic a force-free chromospheric magnetogram, by the so called preprocessing, which was first proposed by \citet{2006Wiegelmann}. We use the preprocessing code developed by \citet{2014Jiang} which is consistent with the CESE-MHD-NLFFF extrapolation code by adopting an optimal magnetic field splitting form. Such a procedure is designed to improve the quality of the raw magnetogram to make it closer to the force-free condition and smooth the raw data to help reduce the measurement uncertainties and numerical errors from the computation. 

\begin{figure*}
\epsscale{0.8}
\plotone{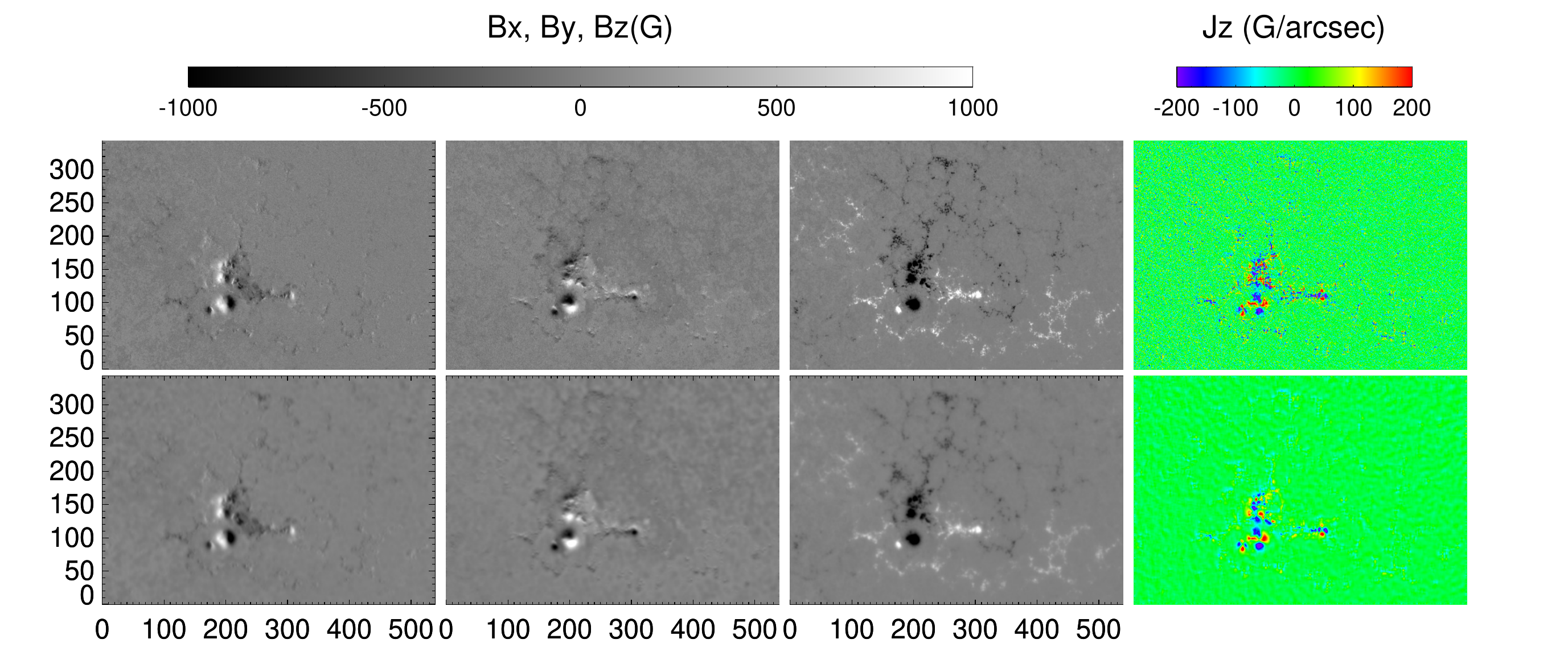}
\caption{First row (from the left to right panels): the three components $B_x,B_y,$ and $B_z$ of the raw magnetogram and the derived vertical current density $J_{z}$ distribution  for event 1, AR11719, at 06:36 UT on 2013 April 11. Second row: the corresponding maps from the preprocessed magnetogram. The size of each map is $540'' \times 344''$. } \label{preprocess1}
\end{figure*}

\begin{figure*}
\epsscale{0.85}
\plotone{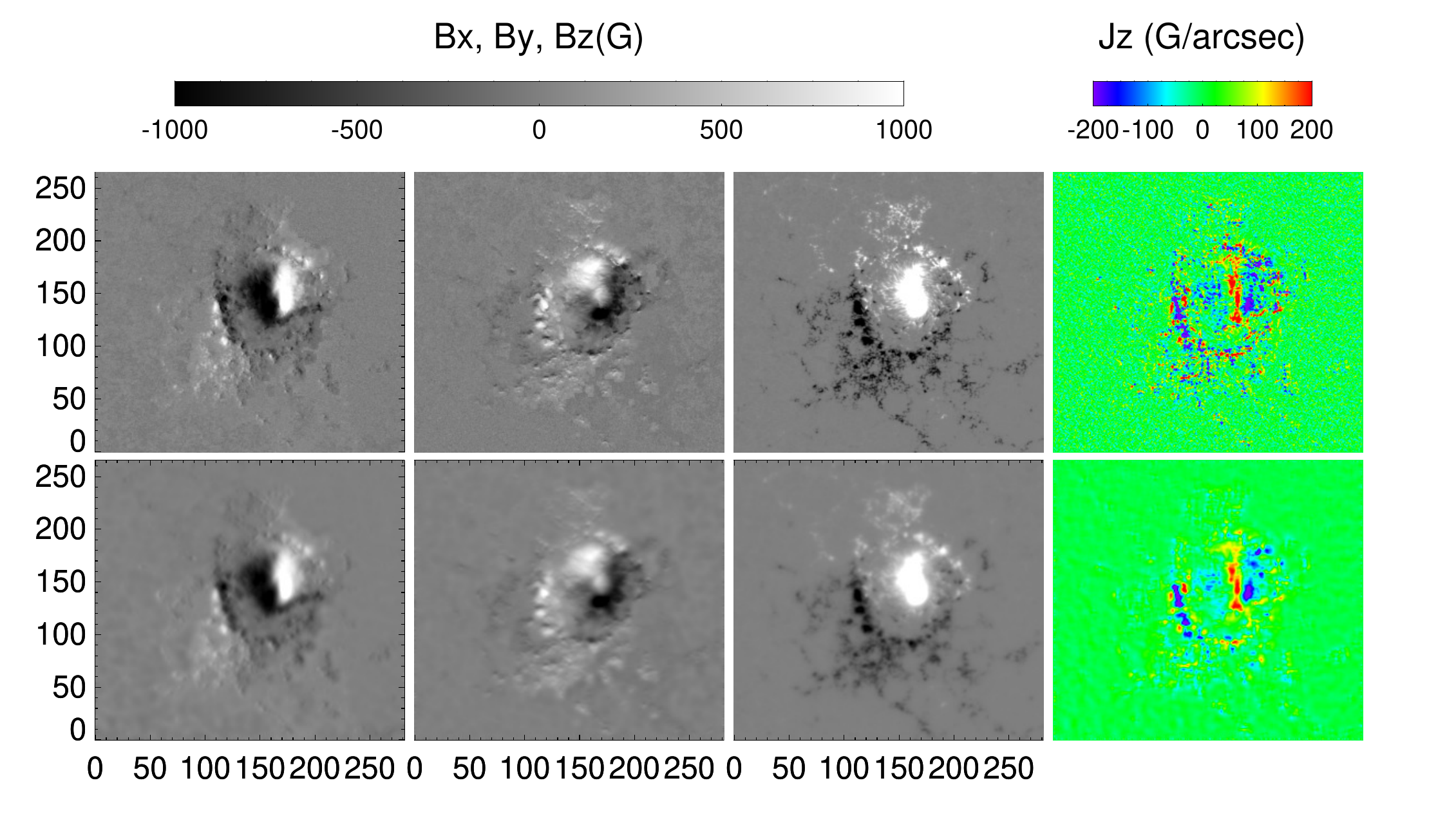}
\caption{The raw and preprocessed maps for event 2, AR 12158, at 17:00 UT on 2014 September 10. Format is the same as Figure \ref{preprocess1}. The size of each map is $282'' \times 266''$. }\label{preprocess2}
\end{figure*}

The high-resolution vector magnetograms are routinely observed by SDO/Helioseismic and Magnetic Imager (HMI; \citealt{2012Schou}). Specifically, the Space-weather HMI Active Region Patches (SHARPs, \citealt{2014Bobra}) vector magnetogram data product, \verb|hmi.sharp_cea_720s|, provided by SDO/HMI, is used as the input for the extrapolation. The SHARP data series provide maps following each patch of significant solar magnetic field for its entire lifetime and the data is also de-rotated to the disk center and remapped using the cylindrical equal area (CEA) Cartesian coordinates. Photospheric vector magnetograms are included with a cadence of 720 s and a spatial resolution of $0.5''$ ($\sim$ 0.36 Mm).  For the two selected events, vector magnetograms from the ARs at least 10 minutes before the flare onset times (estimated from the soft X-ray measurement of the GOES satellite) are preprocessed to get necessary boundary conditions and derive the initial conditions from a potential field solver for the NLFFF extrapolations. The original magnetograms are rebinned from 0.5$''$ per pixel to 1$''$ per pixel for the preprocessing procedure and the subsequent extrapolations. Figures \ref{preprocess1} and \ref{preprocess2} show the overall smoothing effect as a result of preprocessing for the two events by comparing the raw and preprocessed maps of magnetograms and current density $J_{z}$ distributions. Random noise is obviously suppressed  in the $J_{z}$ maps.  

For the consideration of the speed and accuracy of the computation in terms of high-resolution and large field-of-view solar magnetograms, a non-uniform grid structure within a block-structured parallel computation framework is adopted with the help of PARAMESH software package \citep{2000MacNeice} for the CESE-MHD-NLFFF code. For the grid initialization, the whole computation domain includes the pre-set main computation region and the surrounding buffer zones to reduce the influence of the side boundaries \citep{2013Jiang} since the magnetic field at these numerical boundaries are simply fixed as their initial values (i.e., those of the potential field). Then the whole computational domain is divided into blocks with different spatial resolutions and all blocks have  identical logical structures which are evenly distributed among processors. As we vary the grid size only in height (the $z$ dimension) for this study, the grid resolution matches the resolution of the magnetogram at the bottom boundary and decreases by four times at the top of the computational domain. After the grid initialization, the initial condition of the whole computation domain is assigned by a potential field solution derived from the input magnetogram by using the Green's function method \citep{1977Chiu, 2008Metcalf}. 

\subsection{Convergence Study}
Before we start our analysis for the two events, we also examine the relaxation process by several metrics to obtain converged extrapolation results. These include the residual of the field between two successive iteration steps, the force-freeness of the numerical result \textit{CWsin}, the divergence-free condition $\langle|f_{i}|\rangle$ \citep{2000Wheatland, 2008Metcalf} and the total magnetic energy $E_{tot}$ (see their definitions in the Appendix).

\begin{figure*}[htb]
\epsscale{0.85}
\plotone{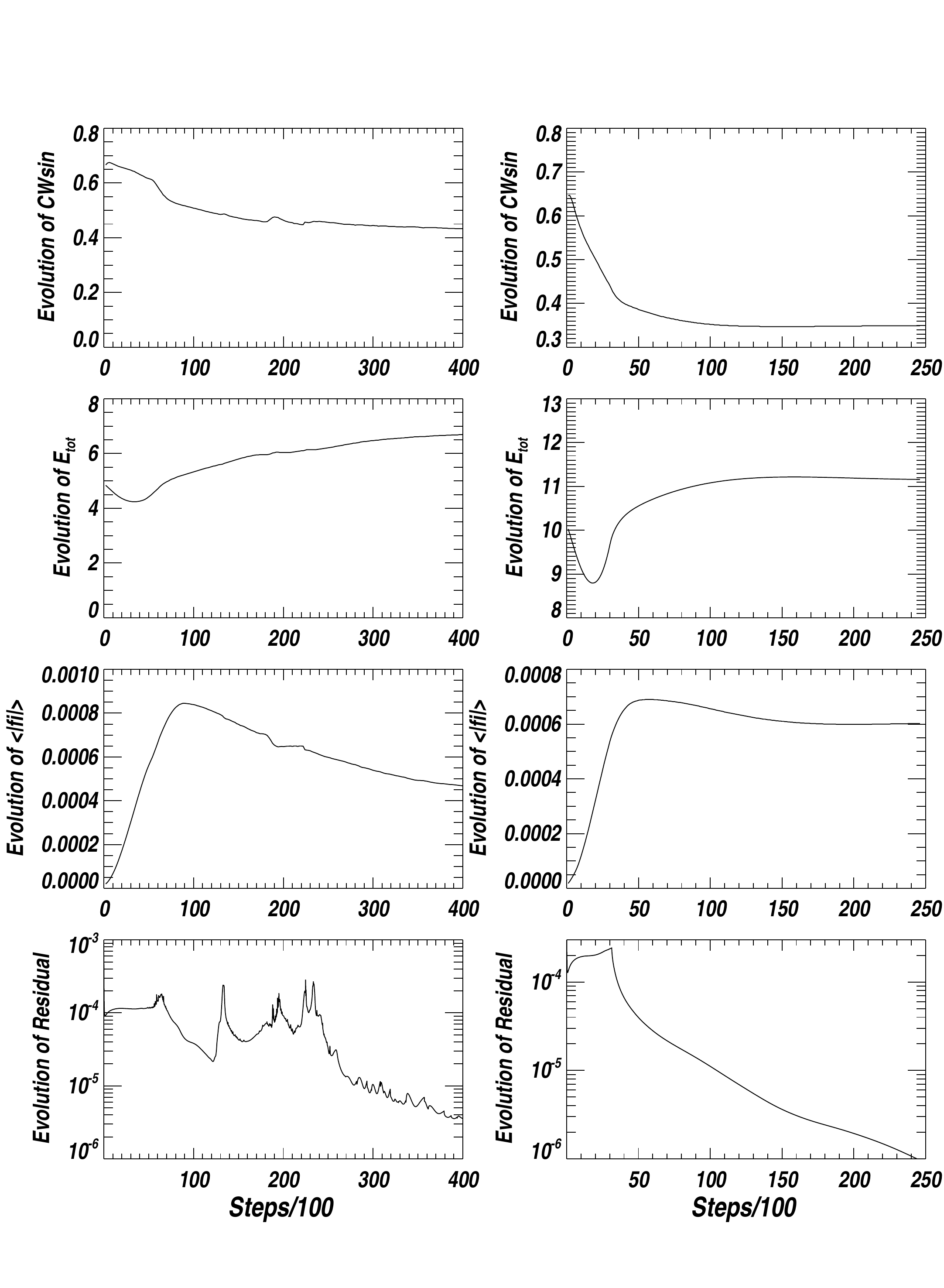}
\caption{Two sets of convergence metrics of two extrapolation runs for AR11719  (event 1, left column) and AR12158  (event 2, right column). The top three rows show the metrics of \textit{CWsin}, the total magnetic energy $E_{tot}$ (arbitrary unit), and $\langle|f_{i}|\rangle$, respectively. Panels in the bottom row show the evolution of residuals for the two events.} \label{converge_all}
\end{figure*}

For event 1, we carry out the extrapolation based on the whole SHARP vector magnetogram ($540'' \times 344''$) and then calculate the convergence metrics for every 200 iteration steps as shown in the first column of Figure \ref{converge_all} with the finest grid size $1''$. As shown in Figure \ref{converge_all}, the residual goes through a gradual increase before $\sim$ 6500 iteration steps because the bottom boundary condition drives the system away from the initial potential field \citep{2012Jiang}. Even though obvious fluctuations appear after the initial driving process, the overall trend of the residual toward the end is decreasing, accompanied by small oscillations. After $\sim$ 30,000 iteration steps, the residual is reduced to $\sim 10^{-5}$ and still maintains a declining trend. Other metrics also display a trend with little variation and both \textit{CWsin} and $\langle|f_{i}|\rangle$ reach relatively small values. Thus the extrapolation result can be considered as a converged solution. It is noticed that there are some oscillations in the convergence process, which may be caused by the broad distribution of the weak field and random noise from the input magnetogram. The total computation time (to converge until 40,000 iteration steps) took about 95 hours with 19 cores on a 24-core local desktop with 48 GB memory.  

For event 2, the size of the SHARP magnetogram is $282''\times 266''$. One run is carried out with the smallest grid size $1''$ and the full size magnetogram. The second column of Figure \ref{converge_all} shows a smooth convergence process. The residual converges very fast after an initial rise exceeding $10^{-4}$ to an order of magnitude smaller, $<$ $10^{-5}$, within 11,000 iterations. All the other metrics show clear monotonic decreases and stabilize after $\sim$ 11,000 iterations, which is consistent with an optimal convergence pattern in the previous tests of this code \citep{2012Jiang, 2012Jiangb, 2013Jiang}. This convergence process is relatively smooth without any spurious oscillations, so a final solution with good indication of convergence is readily obtained for subsequent analysis. It took about 23 hours with 19 cores on the same local desktop for the extrapolation result to converge (after 20,000 iteration steps).

\section{Characterization of MFRs on the Sun}\label{2}
\subsection{MFR Identification Method}
Both extrapolation and observation results are critical for the MFR identification on the Sun. As for the observational analysis of MFRs, we analyze the data from Atmospheric Imaging Assembly (AIA, \citealt{2012Lemen}) and HMI on board the SDO spacecraft to study the evolution of the corresponding flares for the two events. SDO/AIA provides full-disk images in 7 extreme ultraviolet (EUV) and 2 ultraviolet (UV) wavelength channels with a high spatial resolution ($0.6''$ per pixel and a total of 4096 $\times$ 4096 pixels per image) and a moderate time cadence (12 s in EUV channels and 24 s in UV channels). 

To provide additional support for the MFR identification and characterization of the corresponding CME-MFRs at a different stage besides the extrapolation, we also analyze the evolution of flare ribbons and the corresponding magnetic reconnection properties. Flare ribbons map the footpoints of  reconnected field lines. Magnetic reconnection beneath the erupting MFR forms flare loops, and the same amount of reconnected magnetic flux is injected into the MFR. Reconnection may also take place between the erupting MFR and the ambient magnetic field, although this is not the main focus of this study. Therefore, magnetic reconnection flux associated with flare ribbons is useful to establish a quantitative connection between MFRs on the Sun (both before and after the flare eruption) and their interplanetary counterparts. The amount of accumulative magnetic reconnection flux can be measured by summing up the magnetic flux in newly brightened UV pixels within flare ribbons. In this study, we employ 1600 \r{A} data from SDO/AIA and vector magnetograms from SDO/HMI to measure the magnetic reconnection flux and magnetic reconnection rate from the brightening pixels, following the automated approach developed by \citet{2002Qiu,2004Qiu}. The brightening pixels are chosen when the intensity of a pixel is greater than $\sim$ 6 times the median intensity which is fixed and determined from the average of a 6-minute time period (for the region of interest before the eruption).

While an MFR is generally considered as a group of coherent winding field lines, it has not been quantitatively defined in a universal way. Identifying a coherent MFR based on the reconstructed coronal magnetic field derived from the real magnetogram can be difficult, given the complex magnetic topology. \citet{2016LiuR} suggested that the magnetic twist number $T_{w}$ can serve as a good proxy for finding the axis of an MFR. The twist number $T_{w}$ measures how many turns two infinitesimally close field lines wind about each other (see \citealt{2006Berger}), and is defined by
\begin{equation}
\begin{split}
T_{w} &= \int_{L} \frac{\mu_{0} J_{||}}{4\pi B} \,dl = \frac{1}{4\pi}\int_{L} \frac{(\nabla \times \textbf{B})\cdot \textbf{B} }{B^{2}} \,dl, \\
&=\frac{1}{4\pi}\int_{L} \alpha \,dl,  \mbox{ if } \nabla \times \textbf{B} = \alpha \textbf{B}.
\end{split}
\end{equation}
Here $\alpha$ is the force-free parameter and the integral is taken along one magnetic field line with path length $L$, starting from one end point of the field line on the boundary to the other. For both events, extrapolation results are generated utilizing magnetograms that are chosen at least 10 minutes before the flare onset times. We calculate the twist number $T_{w}$ at each grid point in the whole volume with the same grid size as the resolution of the input magnetogram, i.e., 1$''$. Then we start the topology analysis with the definition of \citet{2016LiuR} that an MFR has a bundle of field lines spiraling around the same axis or each other by more than one turn ($|T_{w}| \geqslant 1$, see also \citealt{2019Duan}). Combined with the field-line topology, one may also require that such constrained MFR volume be a single tube without multiple bifurcations. In addition, the footpoints of MFRs should be restricted within or close to main flare ribbon areas identified from AIA observations, given the general relation between the magnetic reconnection process during flares and the formation of erupting CME-MFRs \citep{2001Moore, 2004Qiu, 2009Qiu,2020Zhu}.

\subsection{Results for AR 11719}
\begin{figure}[htb]
\epsscale{1}
\plotone{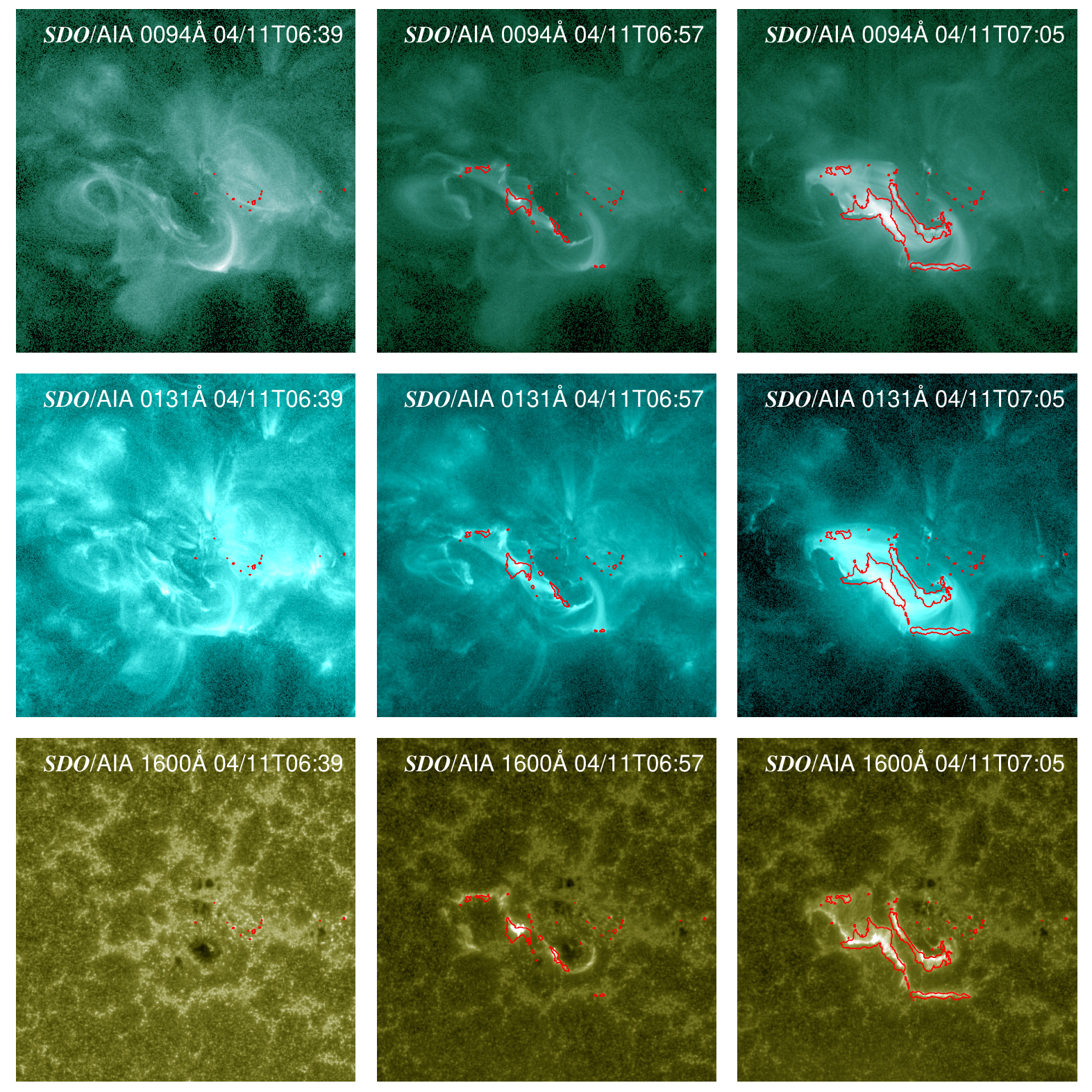}
\caption{Observations from SDO/AIA in 94 \r{A}, 131 \r{A} and 1600 \r{A} (from top to bottom row) wavelength channels at three different times as marked in each panel (from left to right) of AR11719 for event 1. Contours of flare ribbons in red as observed in 1600 \r{A} are also overlaid on 94 \r{A} and 131 \r{A} plots which are observed at the same times.} \label{f0} % pay attention to Dr Qiu's question!
\end{figure}

For event 1 in AR 11719, simultaneous observations of the flare's time evolution in SDO/AIA 94, 131 and 1600 \r{A} wavelength channels before and during the flare eruption are given in Figure \ref{f0}. From the EUV observations in 94 \r{A} and 131 \r{A}, some curved structures are present near the center before the flare eruption, which were recognized as hot channels in \citet{2016Cheng}. But such a sigmoid-like structure based on emission-line images does not necessarily yield a similarly continuous magnetic field-line configuration \citep{1999Titov, 2015Schmieder, 2016Cheng,2017Duan}, i.e., that of an MFR. Instead, these brightened features may correspond to groups of short sheared arcades which are discontinuous, based on the extrapolation result as to be demonstrated below. In the bottom panels of Figure \ref{f0} for the 1600 \r{A} UV observation, there is a typical flare morphology with two brightening ribbons lying nearly in parallel with each other, expanding and then drifting away from each other during the time evolution. The contours of flare ribbons coincide with the curved brightening structures in 131 \r{A} observation at the central region, especially towards the ``hooked'' ends, which gives us a rough estimation of possible positions for the MFR footpoints for this event. 

\begin{figure*}
\begin{center}
\includegraphics[height=12cm]{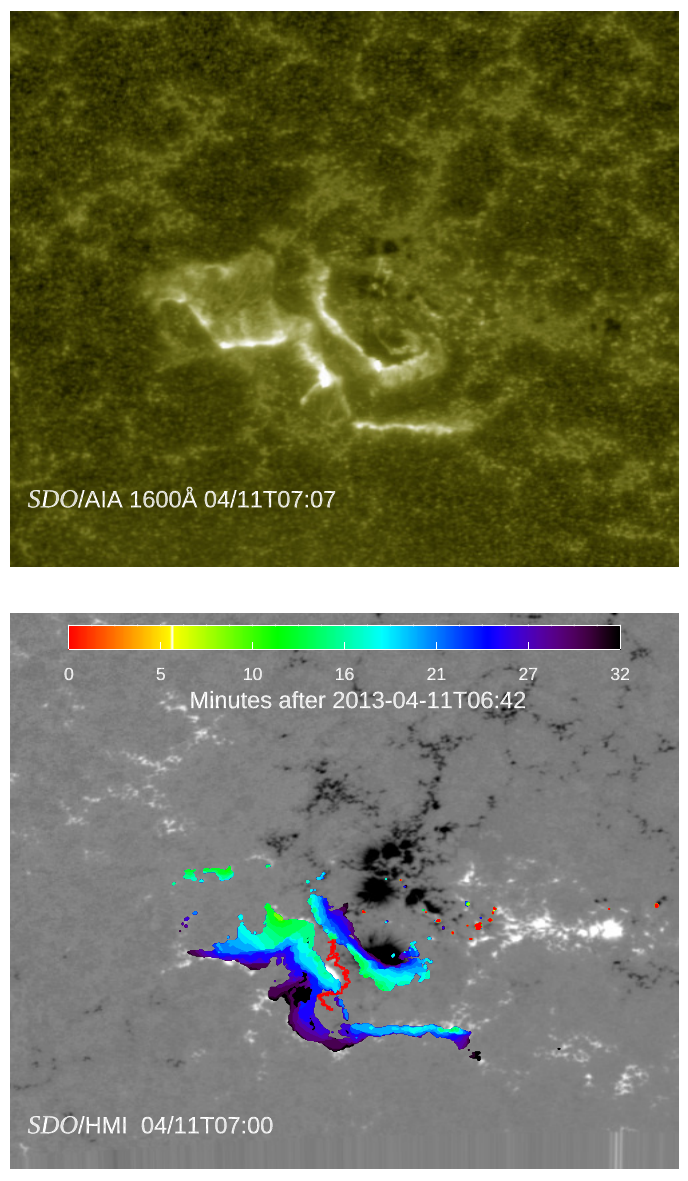}
\includegraphics[height=12cm]{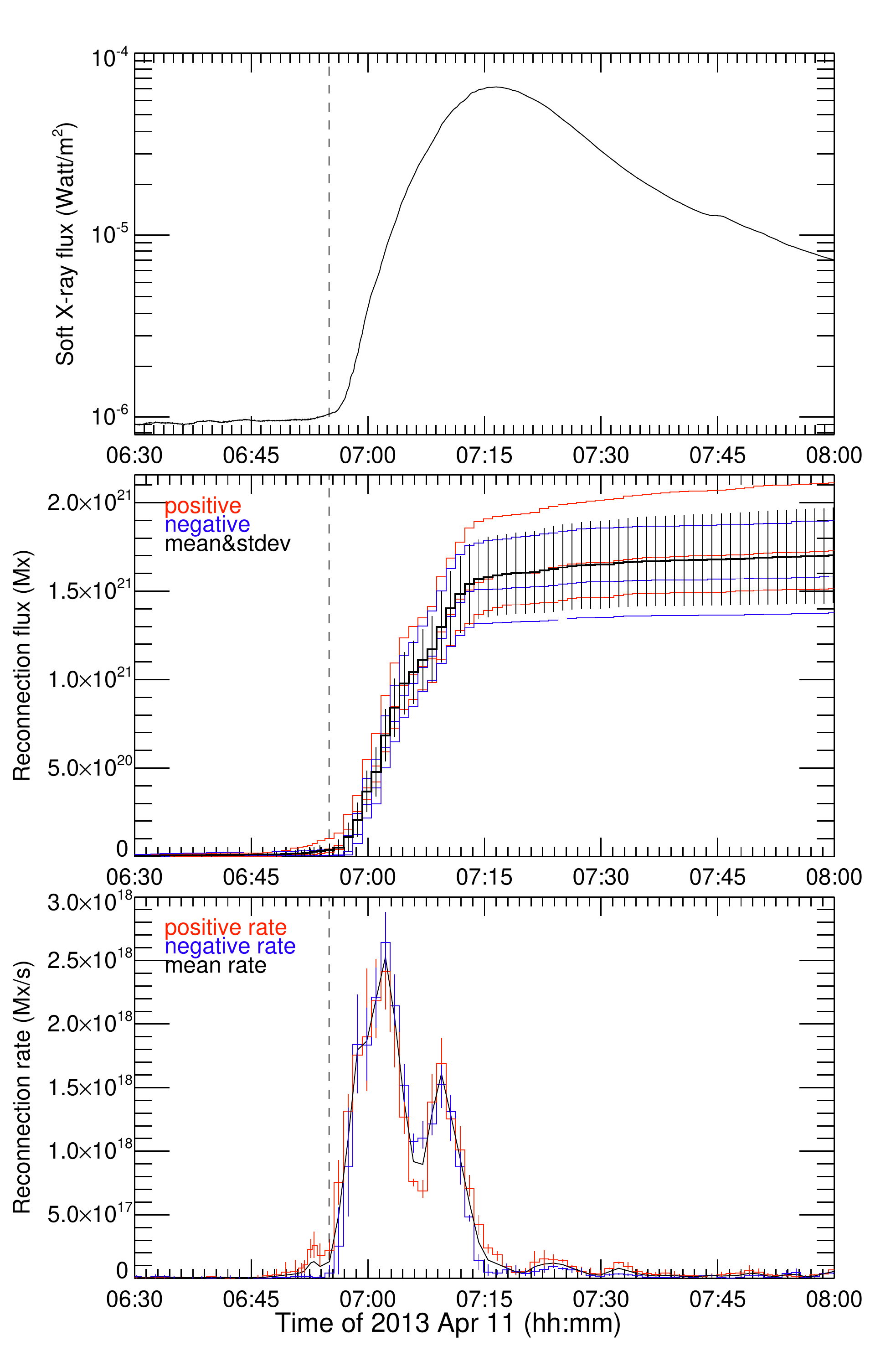}
\end{center}
\caption{Left column: observation of flare ribbons in 1600 \r{A} passband (top panel) and the time evolution of flare ribbons overplotted on the co-aligned HMI magnetogram for event 1 (bottom panel), where the areas swept by flare ribbons are colored by the elapsed time in minutes as denoted by the colorbar. The thick red curve lying in the middle of the lower panel marks the PIL. Right column: GOES soft X-ray (1-8 \r{A}) flux measurement for the flare in event 1 (top panel), the magnetic reconnection flux measured from 1600 \r{A} observation (middle panel) for both positive (red) and negative (blue) flux measurements with uncertainty limits based on different background removal criteria, while the unsigned mean flux is shown in black with the standard deviation represented by the errorbars, and the corresponding magnetic reconnection rates for event 1 (bottom panel). The dashed lines in all three panels indicate the flare onset time of event 1.} \label{f2}
\end{figure*}

The time evolution of the flare ribbons in the corresponding SDO/HMI magnetograms which are remapped to the sub areas in SDO/AIA's field of view is shown in Figure \ref{f2}, the left column. Besides, we add the X-ray flux measurement of the whole Sun provided by the GOES satellite for the wavelengths of soft X-ray (1 - 8 \r{A}) during the same time period in the right column, together with the concurrent measurements of accumulative magnetic reconnection flux and magnetic reconnection rate by the approach of \citet{2002Qiu}. This M6.5 flare eruption starts at $\sim$ 06:55 UT according to the rapid change of the soft X-ray flux curve, which is consistent with the onset of the magnetic reconnection flux increase shown in the second panel in the right column. Based on the average of the total unsigned magnetic flux in each enclosed ribbon area with one dominant polarity (either positive or negative,  \citealt{2017Kazachenko}), the final accumulative magnetic reconnection flux reaches the magnitude of $17 \pm 2.8 \times 10^{20}$ Mx after the eruption. Given the association between the magnetic reconnection flux and the flux content of the corresponding ICME/MC flux ropes \citep[e.g.,][]{2007Qiu,2014Hu}, such flux measurement from flare ribbons can be helpful for making further connections of MFRs on the Sun and their in situ counterparts, as to be laid out in Section \ref{3}. 

\begin{figure*}
\gridline{\fig{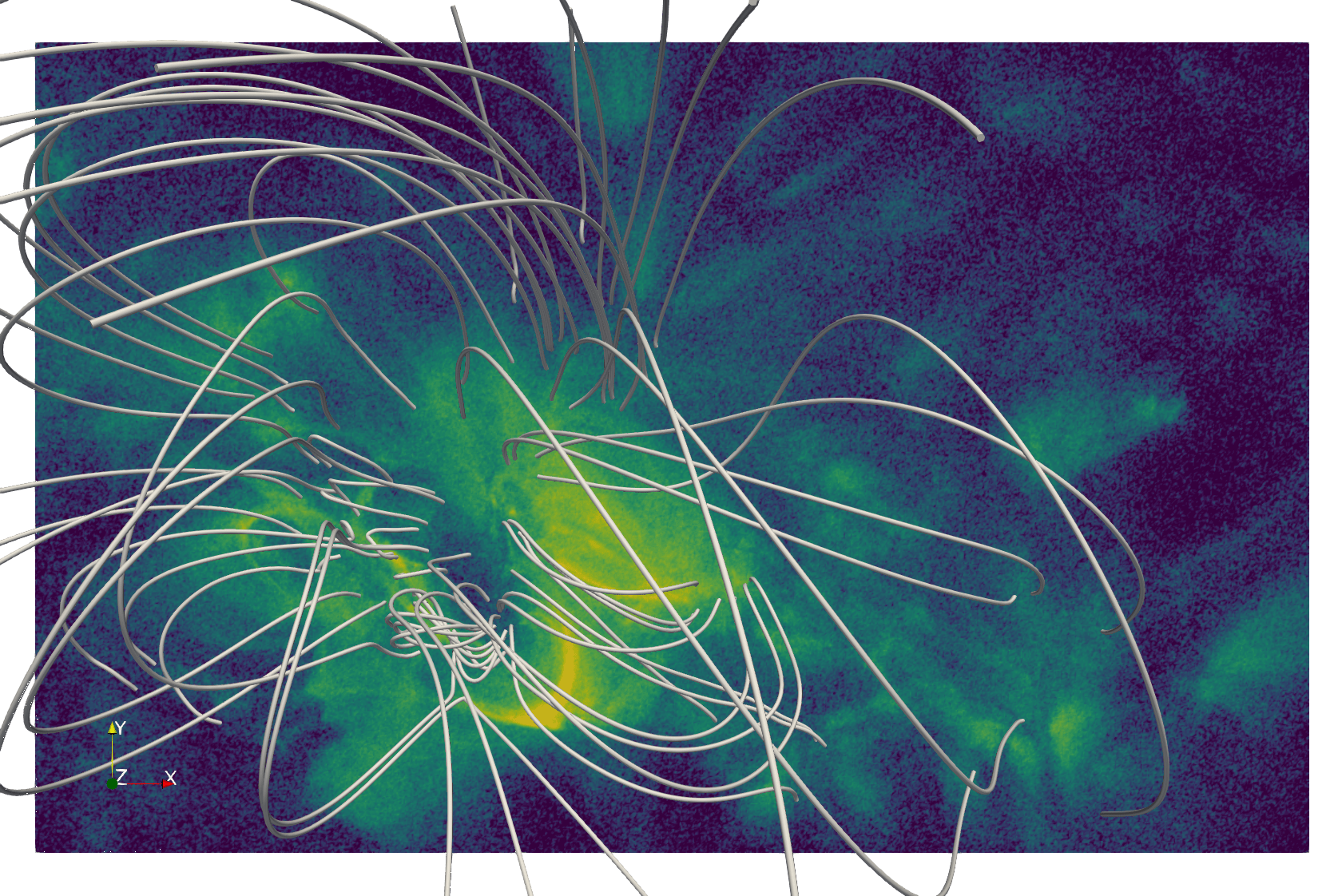}{0.45\textwidth}{(a)}
           \fig{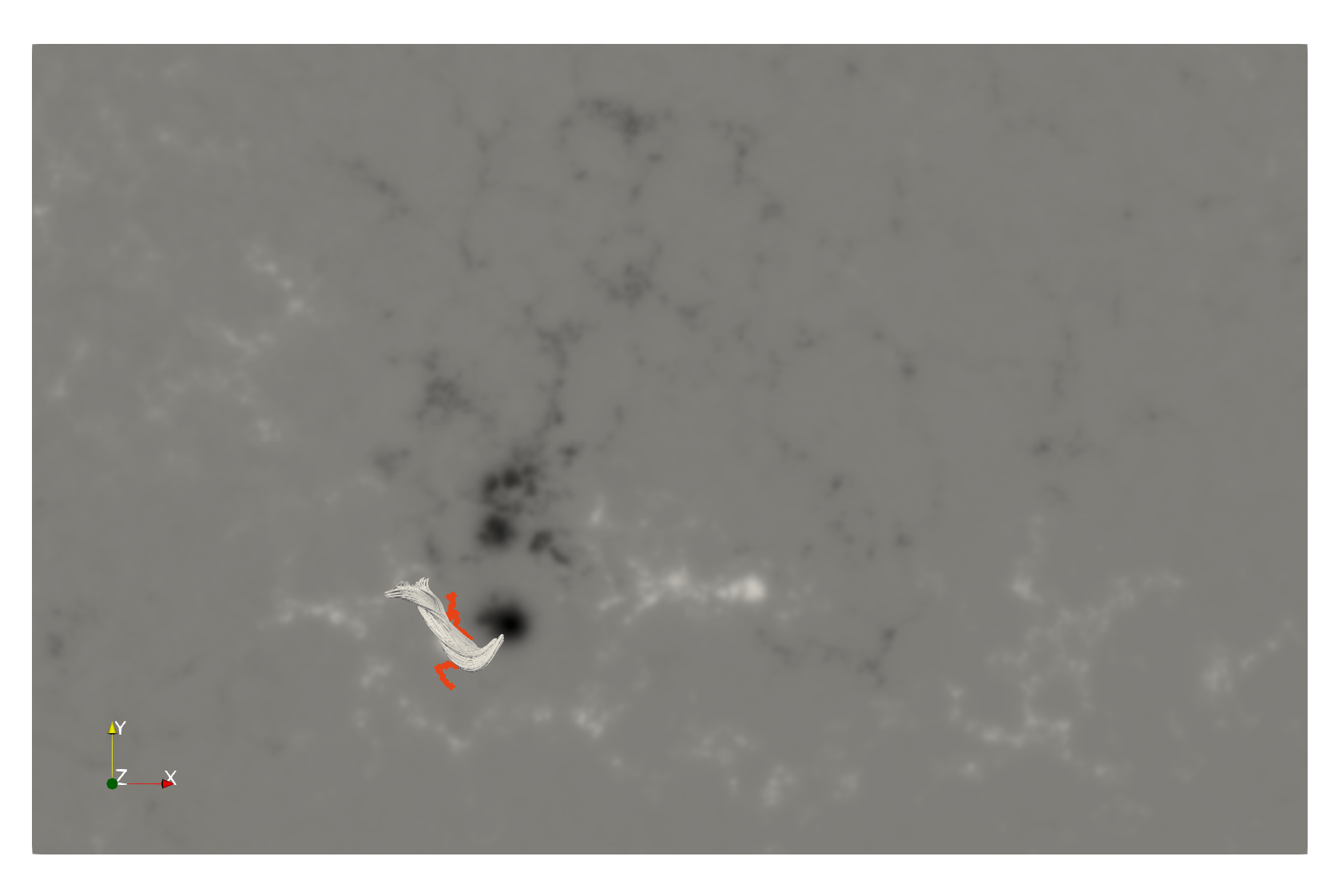}{0.45\textwidth}{(c)}
          }
\gridline{\fig{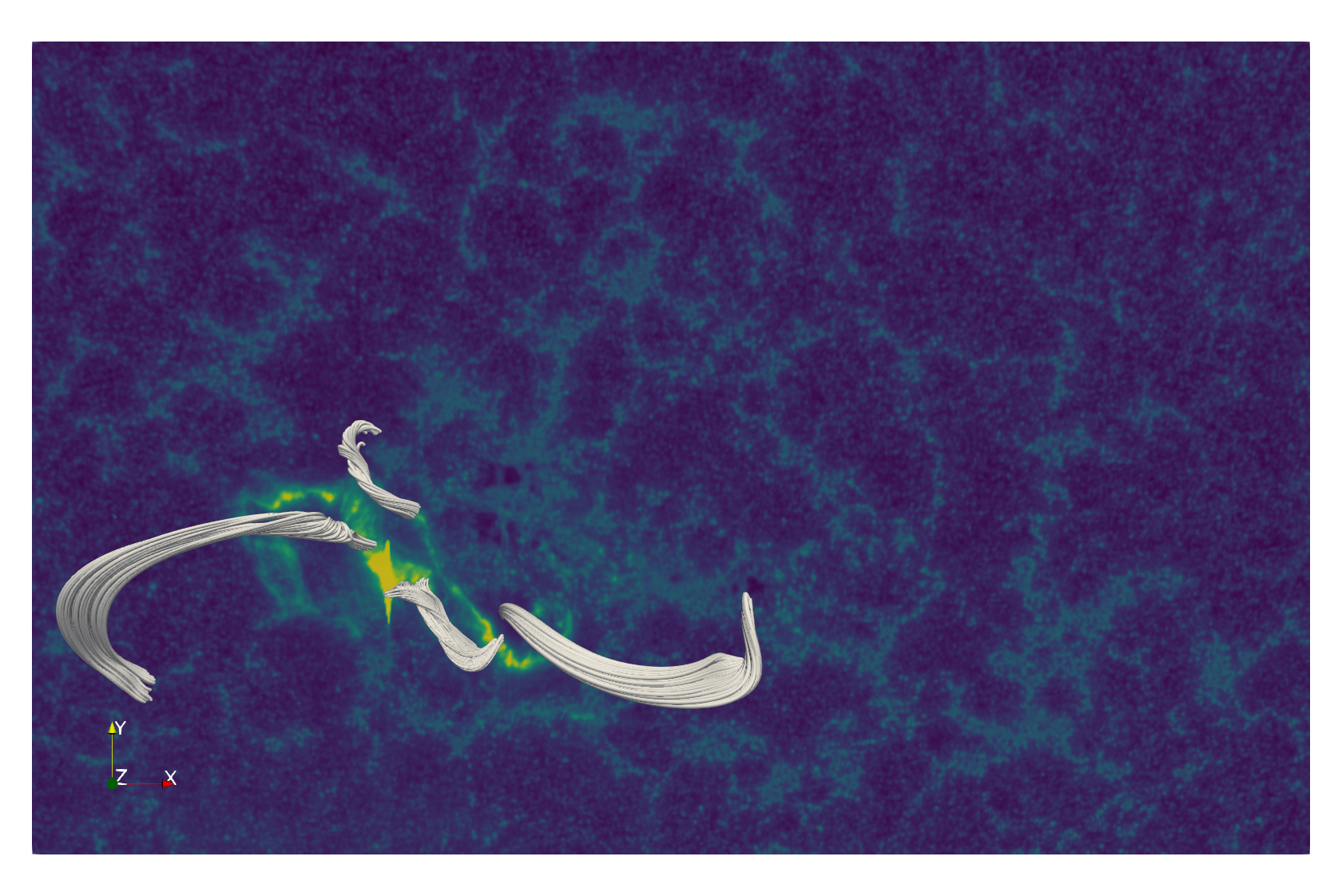}{0.45\textwidth}{(b)}
           \fig{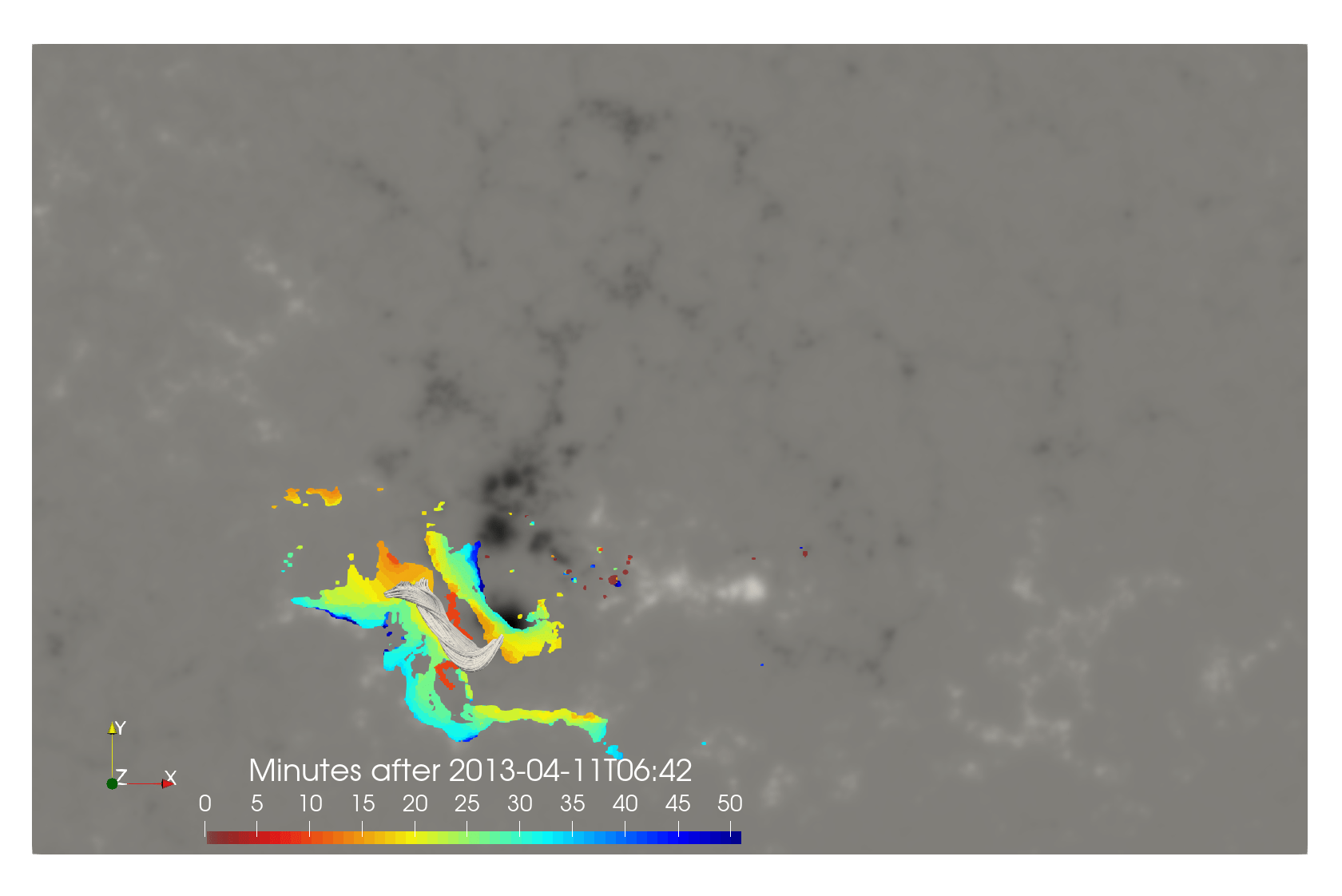}{0.45\textwidth}{(d)}\textbf{}
          }
\caption{Identification of an MFR for event 1: (a) the overall field line configuration superimposed on an AIA 94 \r{A} image which is observed at 06:36 UT, (b) field line groups identified from the extrapolation result based on the criterion of $|T_{w}| > 1$ overplotted on an AIA 1600 \r{A} image observed at 06:59 UT, (c) the field lines of the identified MFR,  and the underlying PIL in red drawn over the corresponding line-of-sight HMI magnetogram, and  (d)  the same as (c) except for the additional superimposed flare ribbons, which are color coded by elapsed time in the same way as Figure \ref{f2}.} \label{f11}
\end{figure*}

In Figure \ref{f11}a, some sample field lines are drawn over the corresponding AIA 94 \r{A} image to give a qualitative comparison between the extrapolation result and the observation. Several loop structures are recovered overlapping with selected field lines, and a set of twisted field lines lying around the PIL takes the shape resembling the middle of the inverse ``S'' sigmoid as seen in 94 \r{A} channel (see also Figure \ref{f11}c). Among the comparisons with the 1600 \r{A} observation in Figure \ref{f11}b, footpoints of the twisted field lines locating close to the flare ribbons are associated with grid points with negative $T_{w}$ values. On a plane near the bottom layer (at $z$ = 2$''$ above the photosphere), we pick all the points with $T_{w} \leqslant -1$  around the central sigmoidal structure, and plot field lines passing through this set of seed points. We eliminate open field lines which only have one end point attached to the bottom boundary thus not ``closed'', and also ill-defined $T_{w}$ values. As a result shown in Figure \ref{f11}b, four groups of field lines are distinguished starting with the selected seed points. Compared to the locations of the flare ribbons, three groups of field lines are excluded since a part of their footpoints extends out of the ribbon sites. Therefore the remaining bundle of field lines shown in Figure \ref{f11}c is identified to be the most likely candidate MFR for the 2013 April 11 event before the flare eruption. After determining the MFR, we can find the axial field line with the maximum $|T_{w}|$ of the MFR. The axis of the identified MFR in event 1 possesses $T_{w} = -1.5$ which lies close to the bottom boundary and reaches a maximum height at $z \sim$ 19$''$. The time sequence of flare ribbons after 6:42 UT is then co-aligned with the bottom boundary magnetogram and overplotted in the usual way, color coded by elapsed time in Figure \ref{f11}d. It shows that two groups of footpoints from the identified MFR locate on the opposite sides of the flare ribbons near the far ends, consistent with the scenario proposed by \citet{2001Moore}. 

\begin{figure}[ht]
 \gridline{\fig{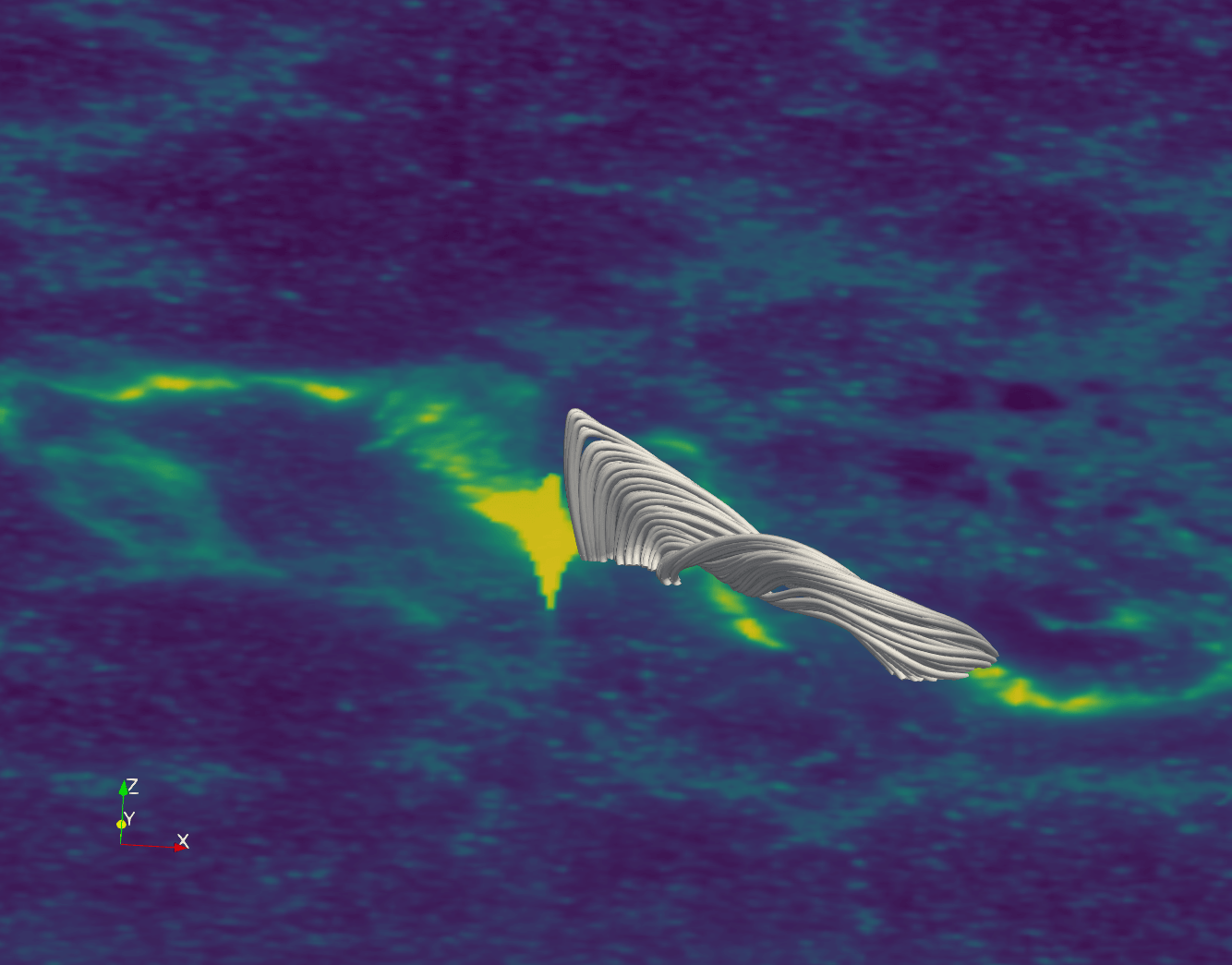}{0.3\textwidth}{(a)}
           \fig{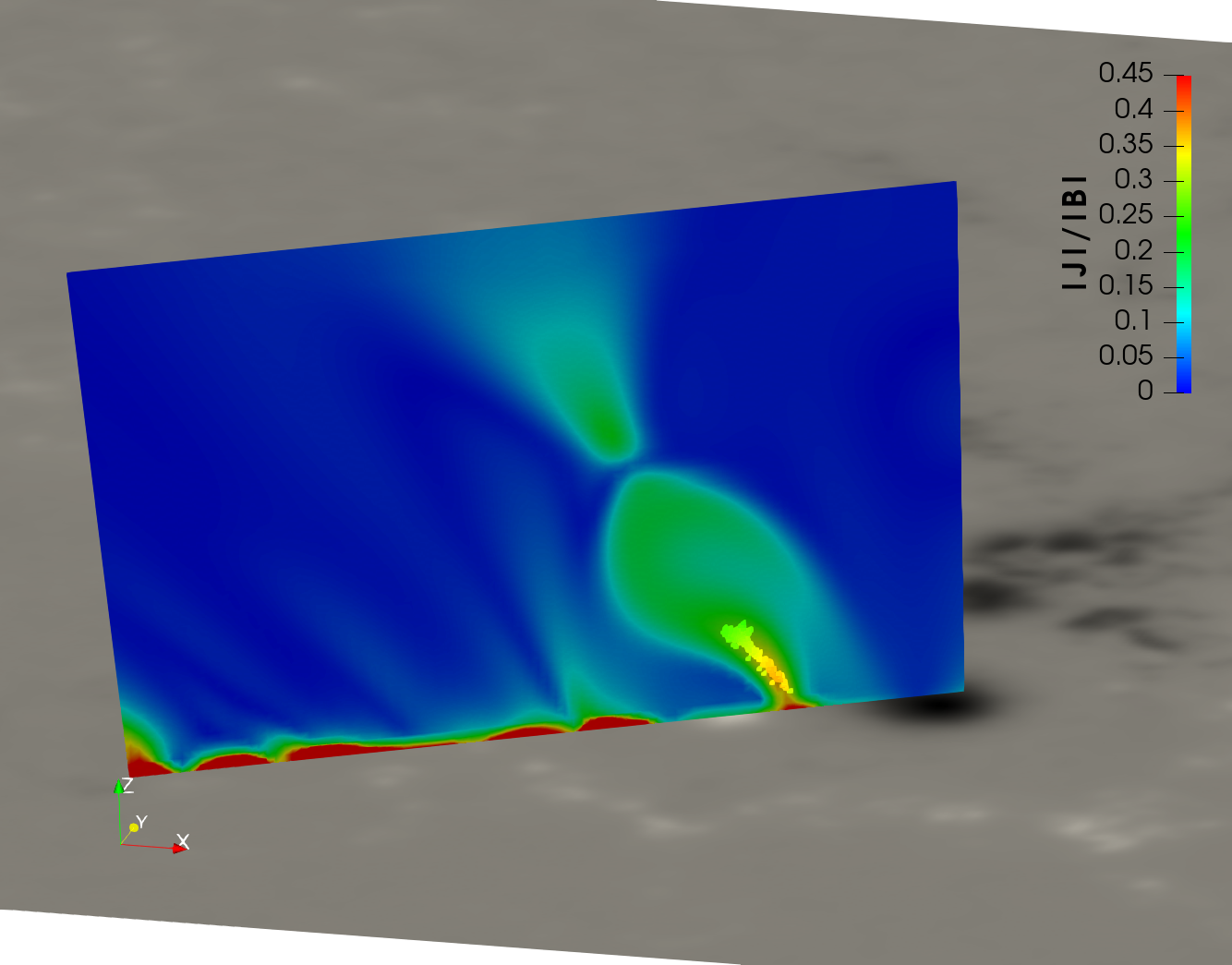}{0.3\textwidth}{(c)}
           \fig{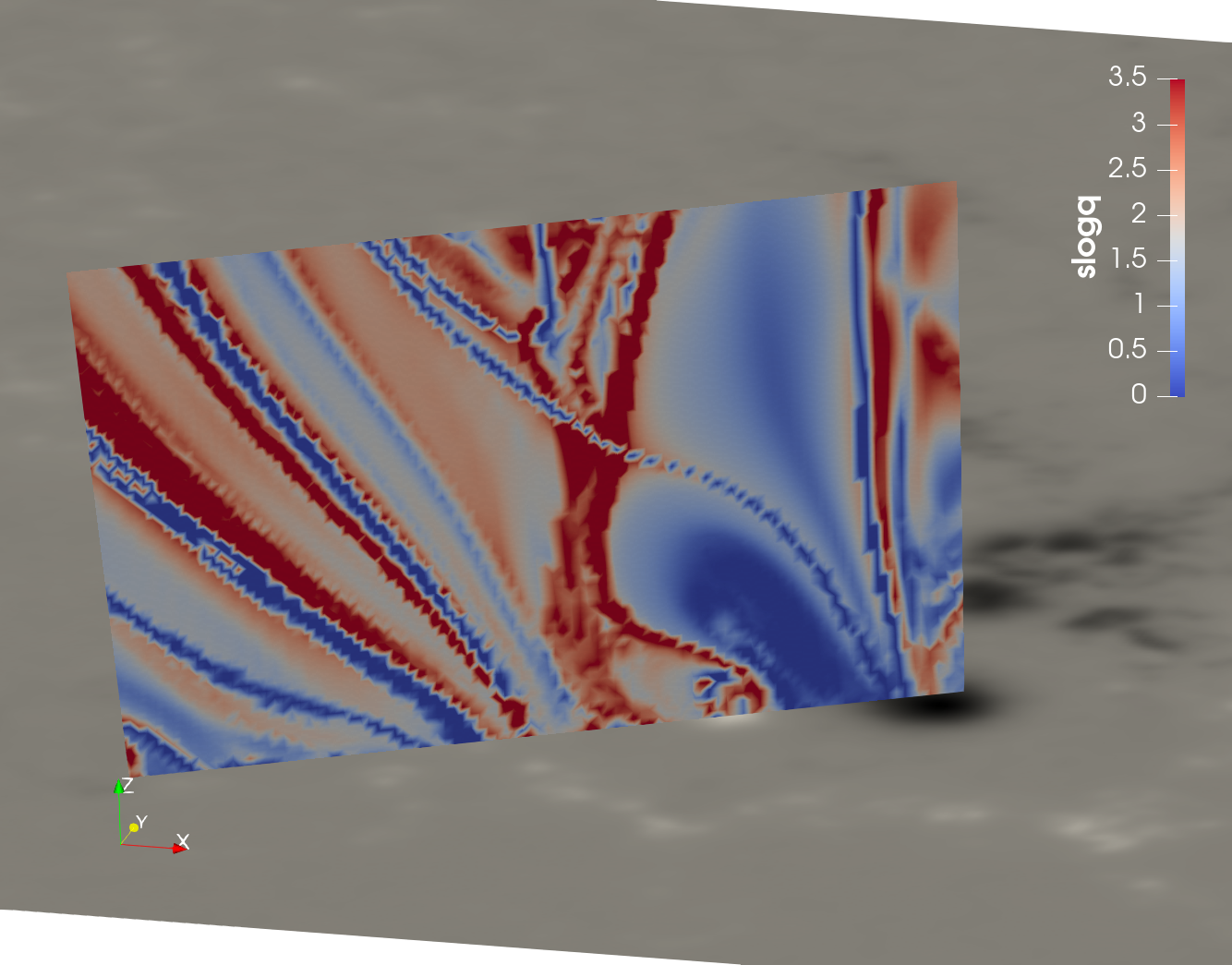}{0.3\textwidth}{(e)}
          }
 \gridline{\fig{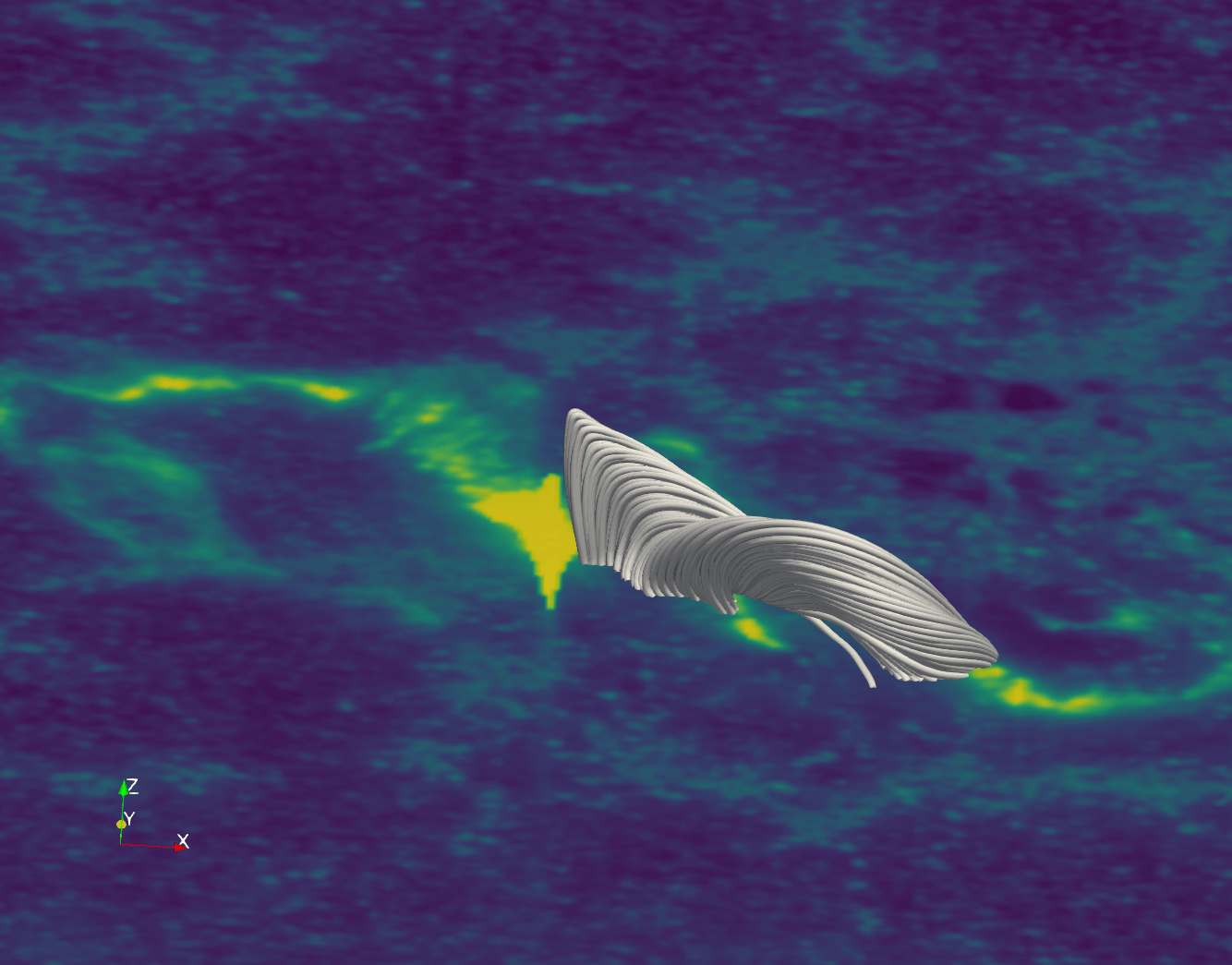}{0.3\textwidth}{(b)}
           \fig{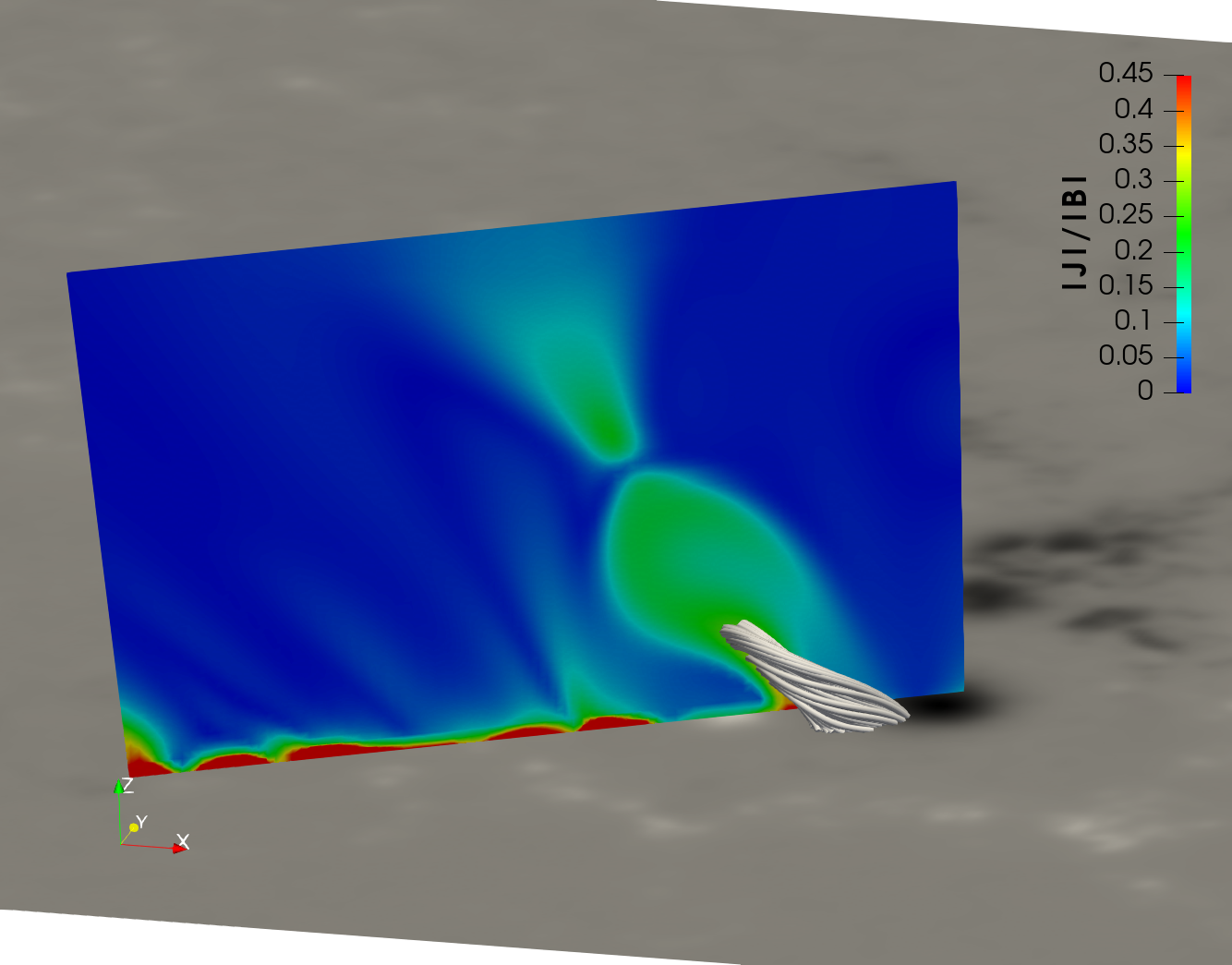}{0.3\textwidth}{(d)}
           \fig{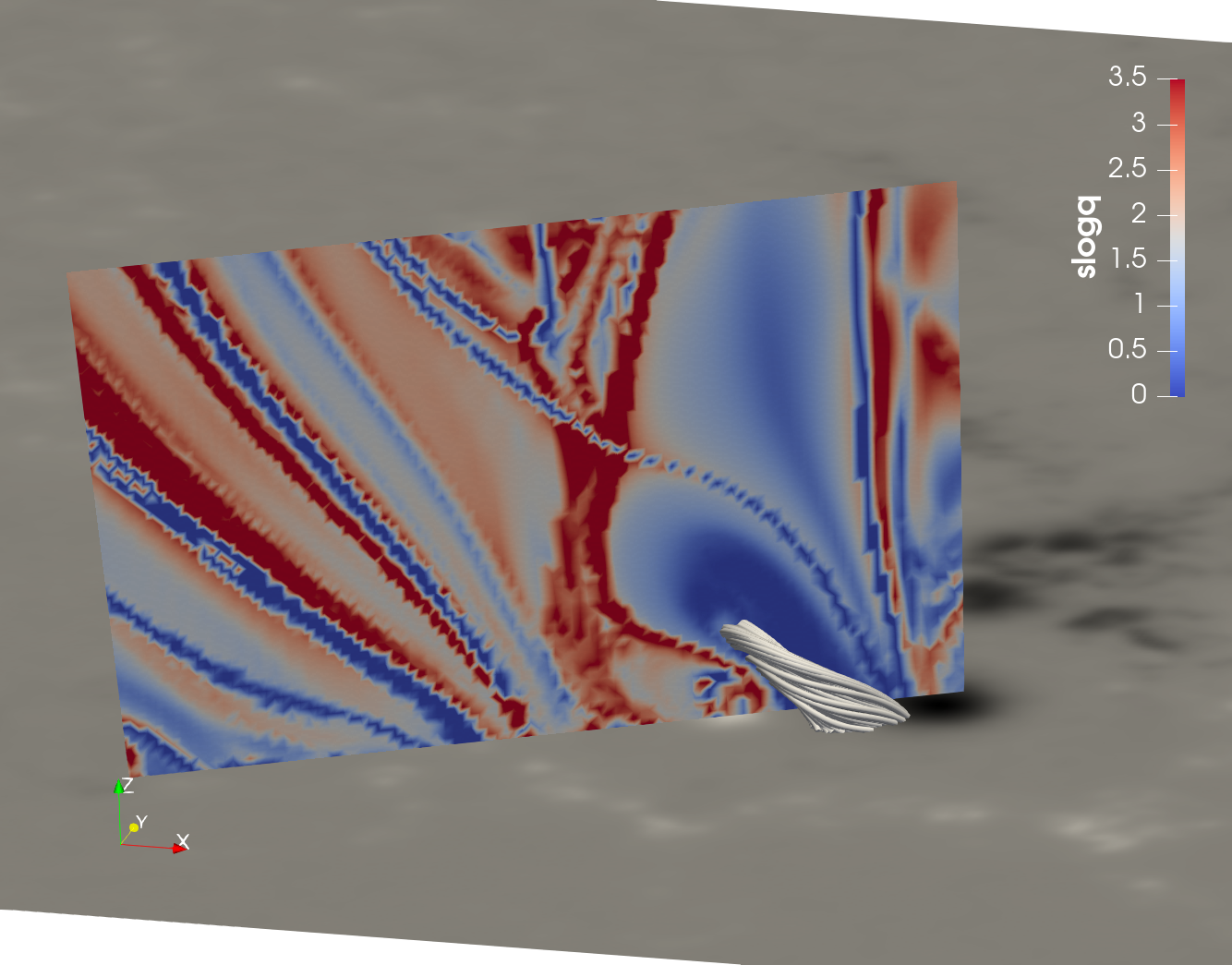}{0.3\textwidth}{(f)}
          }
  \caption{The enlarged and  side views of the identified MFR in Figure \ref{f11}b for the criteria of (a) $|T_{w}| > 1$, and (b) $|T_{w}| > 0.8$, respectively, and similarly for Figure~\ref{f11}c:  (c) the distribution of $|\mathbf{J}|/|\mathbf{B}|$ as indicated by the colorbar on a vertical slice across the identified MFR for $|T_{w}| > 1$ with the MFR field-line intersection points colored by the corresponding values according to the colorbar, (d) same as (c) but overplotted with the MFR field lines, and (e-f) the distributions of the squashing degree Q on the same slices as (c) and (d).} \label{f12}
\end{figure}

To further confirm the existence of the MFR, we also check different topological properties from a side view. In Figure \ref{f12}, the coherent MFR structure is still maintained with a different $|T_{w}|$ threshold as seen in Figure \ref{f12}a and \ref{f12}b.   The distribution of the quantity $|\textbf{J}|/|\textbf{B}|$ as a proxy to current density  is displayed in Figure \ref{f12}c and \ref{f12}d  on a cross section plane nearly perpendicular to the MFR. The current density $|\mathbf{J}|$ itself also shows a similar distribution.  Based on the current density distribution at the intersections between the identified MFR field lines and the vertical slice in Figure \ref{f12}c, the flux rope goes through a region with relatively high current density. The geometric boundary of an MFR can also be estimated by the location of a quasi-separatrix layer (QSL), a very thin layer where there is a strong gradient of the field line connectivity \citep{1996Demoulin}. Such a feature is typically defined mathematically by the high squashing factor Q \citep{2002Titov, 2021Vemareddy}. As shown in Figure \ref{f12}e and \ref{f12}f, the identified group of field lines with small Q values is surrounded indeed by a clear boundary with high squashing degree Q. 

\subsection{Results for AR 12158}
\begin{figure}[htb]
\epsscale{1}
\plotone{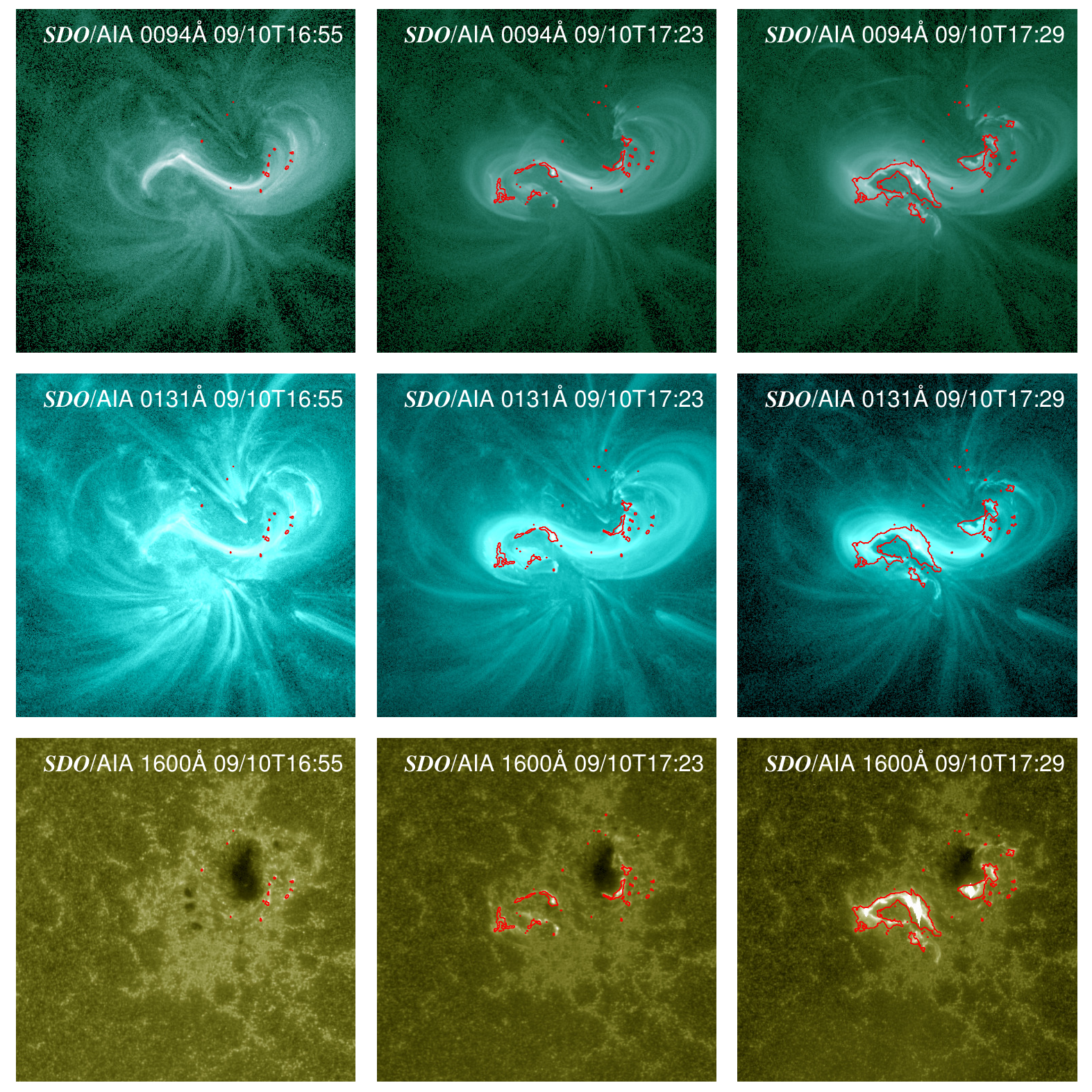}
\caption{Observations from SDO/AIA in 94 \r{A}, 131 \r{A} and 1600 \r{A} (from top to bottom row) wavelength channels of AR12158 for event 2. The format is the same as Figure~\ref{f0}.} \label{f1}
\end{figure}

Observations of the flare ribbon evolution before and during the flare eruption for event 2 in AR 12158 are shown in Figure \ref{f1}. Event 2 also exhibits a two-ribbon flare morphology, with two ribbon areas co-located near the two ends of an inverse ``S'' shape sigmoidal structure. The southward ribbon has a more dominant swept area in size than the other. Similarly, we use the overlapping regions between the flare ribbon areas and the curved brightening sigmoidal structures in 131 \r{A} observation to approximate the possible locations of MFR footpoints in this event. 

\begin{figure}[htb]
\begin{center}
\includegraphics[height=12cm]{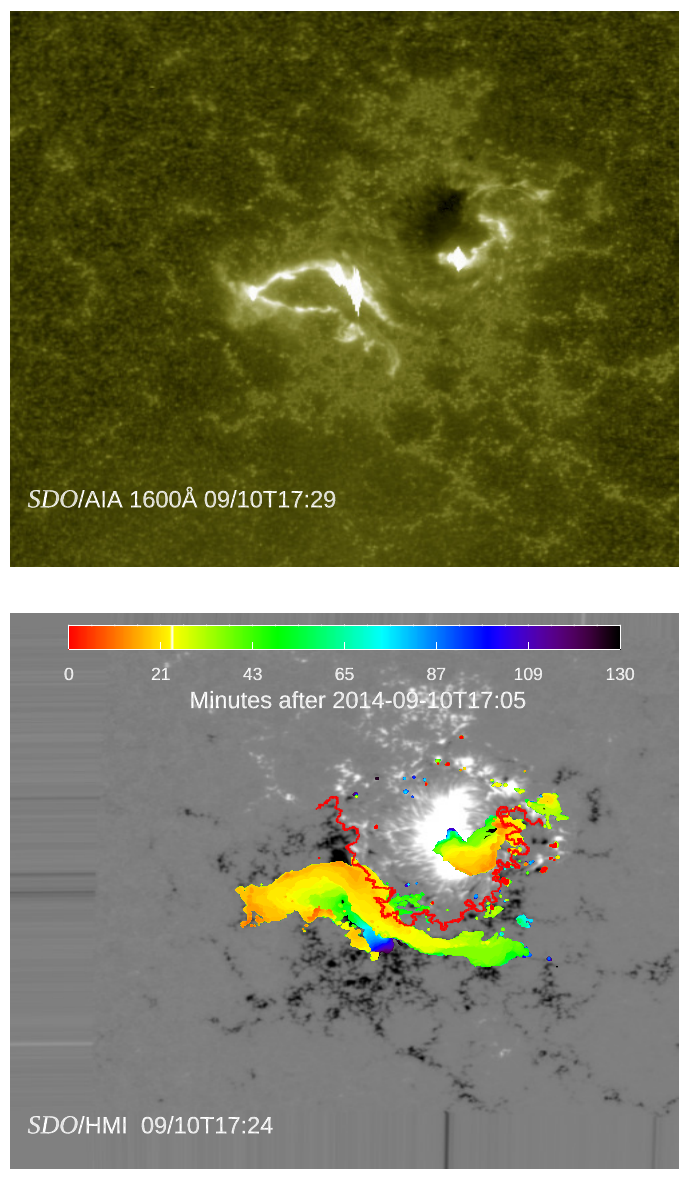}
\includegraphics[height=12cm]{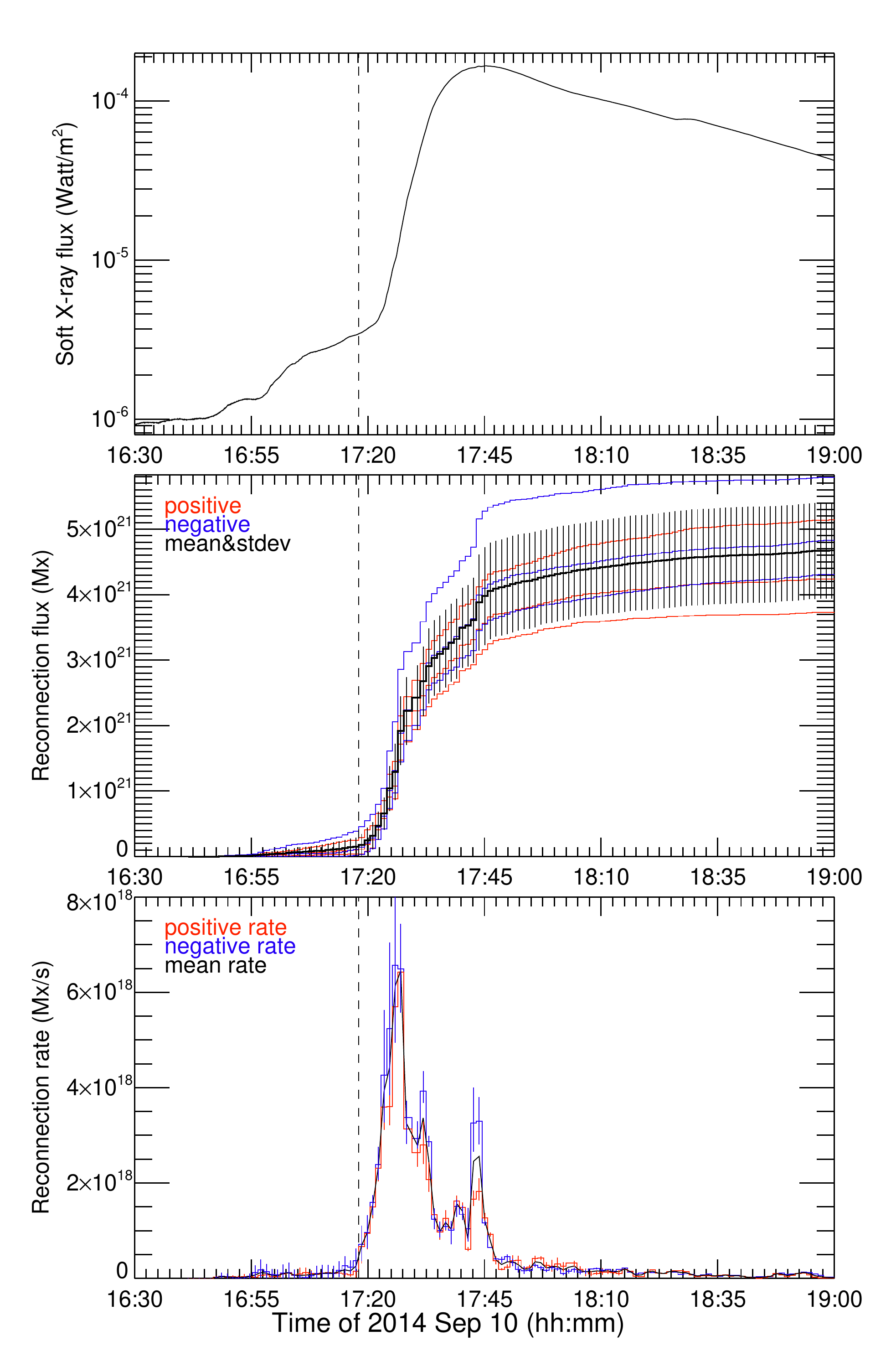}
\end{center}
\caption{The observations of flare ribbons and measurements of the reconnection flux for event 2. Format is the same as Figure \ref{f2}.
} \label{f3}
\end{figure}

In Figure \ref{f3}, we show the same set of panels for event 2, as in Figure \ref{f2}. We find that the initial enhancement of the X-ray flux is earlier than the significant increase of the magnetic reconnection flux. After a slow rise phase with small reconnection rate, a strong flare is produced quickly after $\sim$17:20 UT. The final accumulative magnetic reconnection flux reaches the magnitude of $47\pm 7.5 \times 10^{20}$ Mx for event 2. The flare ribbon morphology still exhibits general features for a ``two ribbon'' flare, albeit asymmetry of the spatial distributions is more pronounced, indicating perhaps more significant deviation from a ``standard'' 2D geometry. 

\begin{figure}%[htbp]
\gridline{\fig{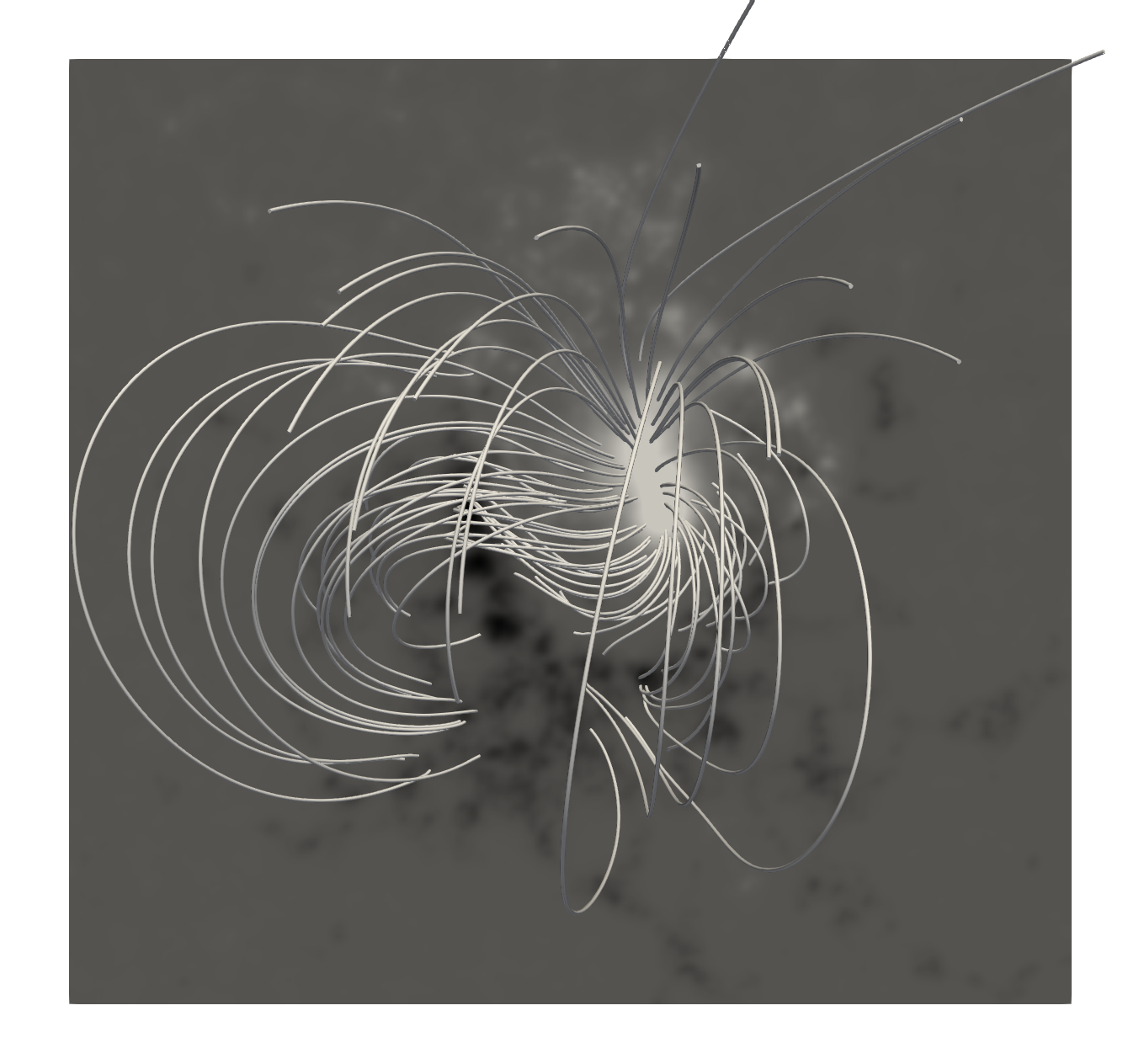}{0.3\textwidth}{(a)}
           \fig{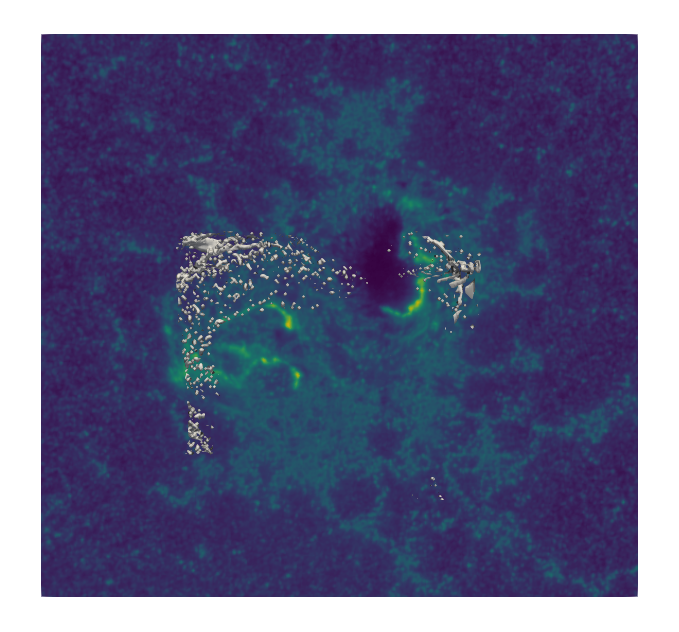}{0.3\textwidth}{(c)}
           \fig{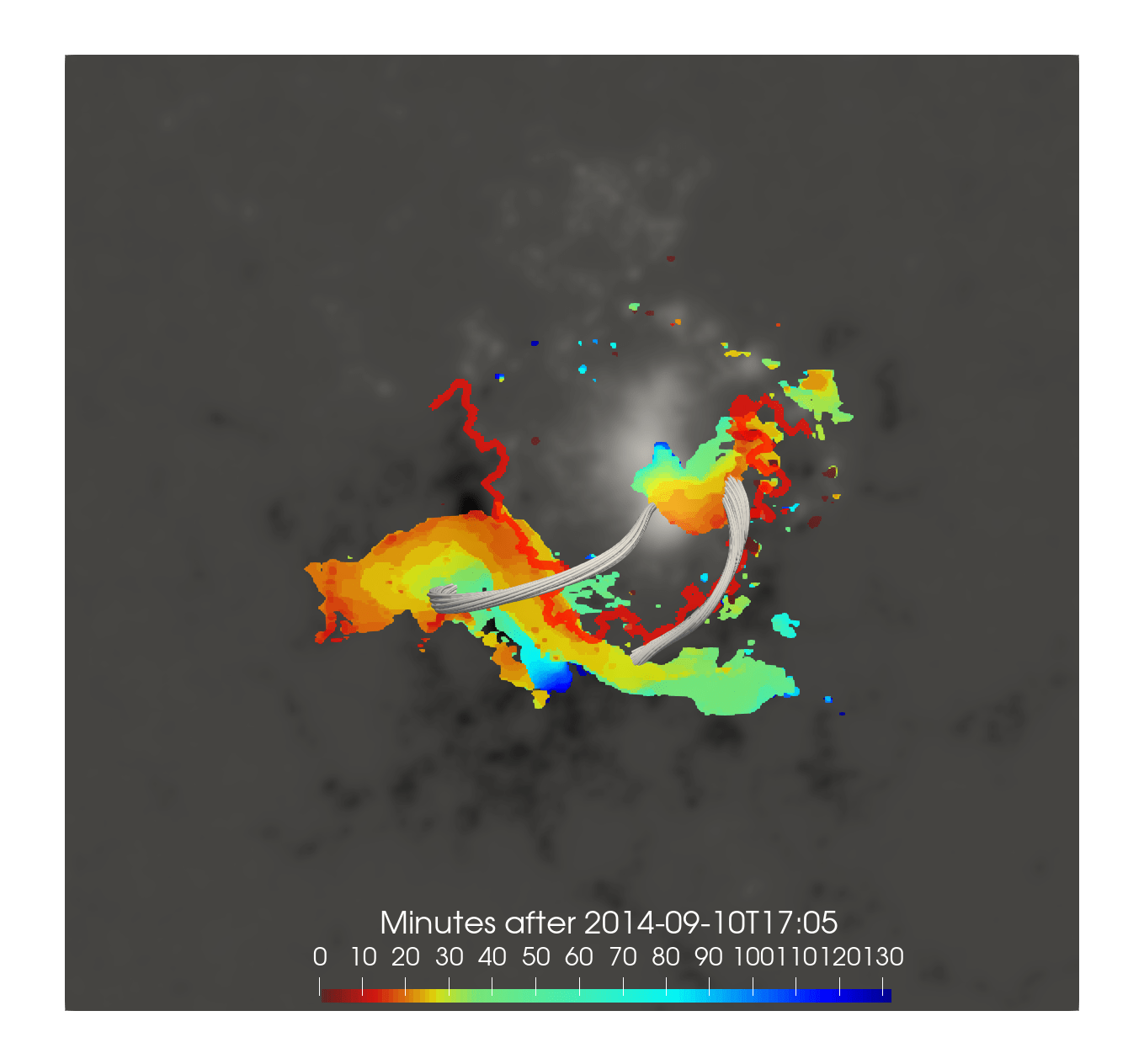}{0.3\textwidth}{(e)}
          }
\gridline{\fig{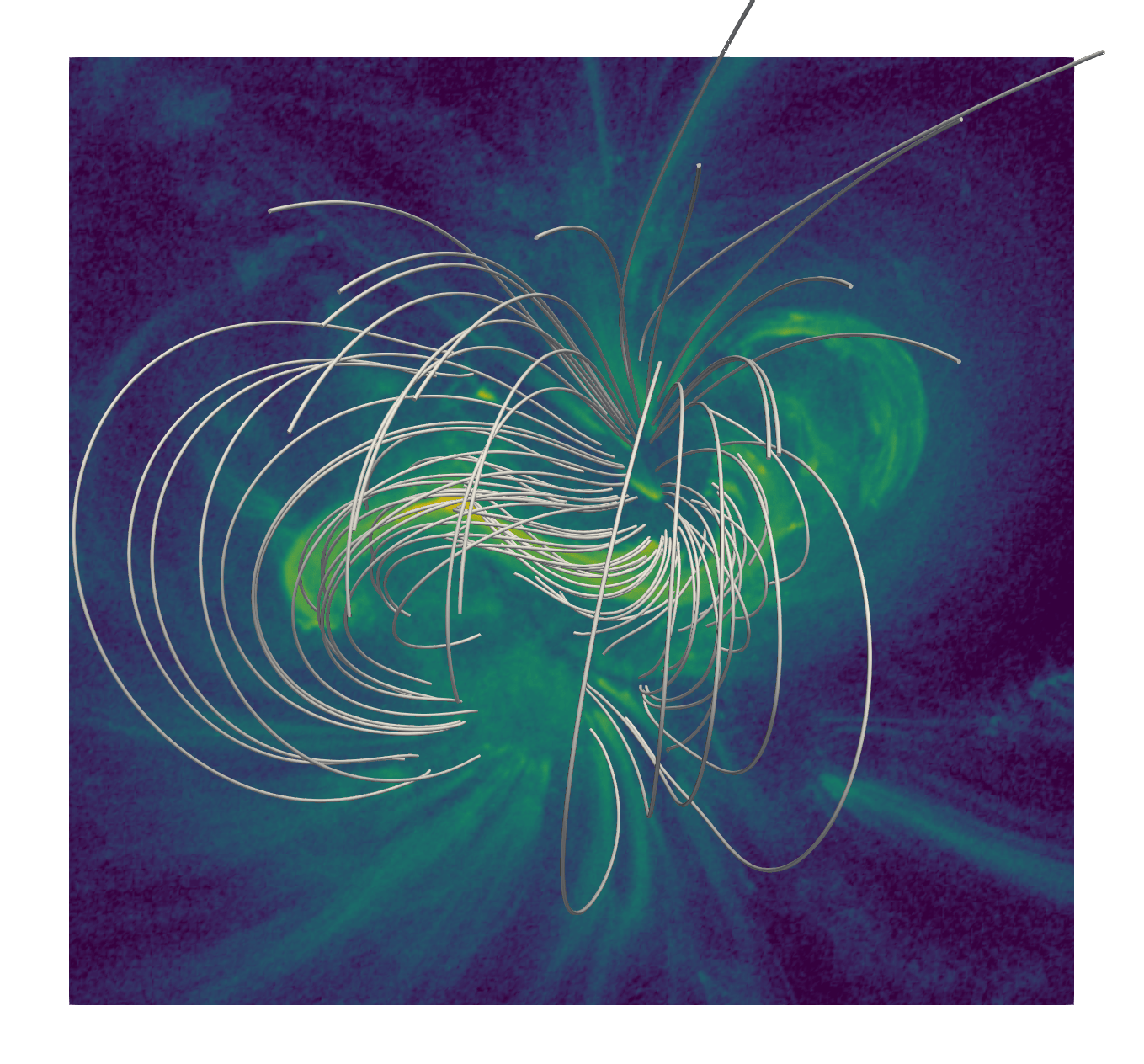}{0.3\textwidth}{(b)}
           \fig{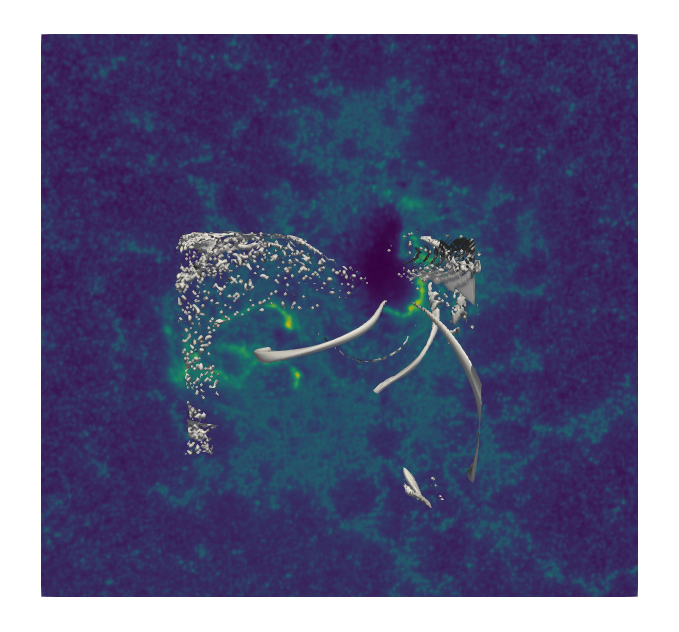}{0.3\textwidth}{(d)}
           \fig{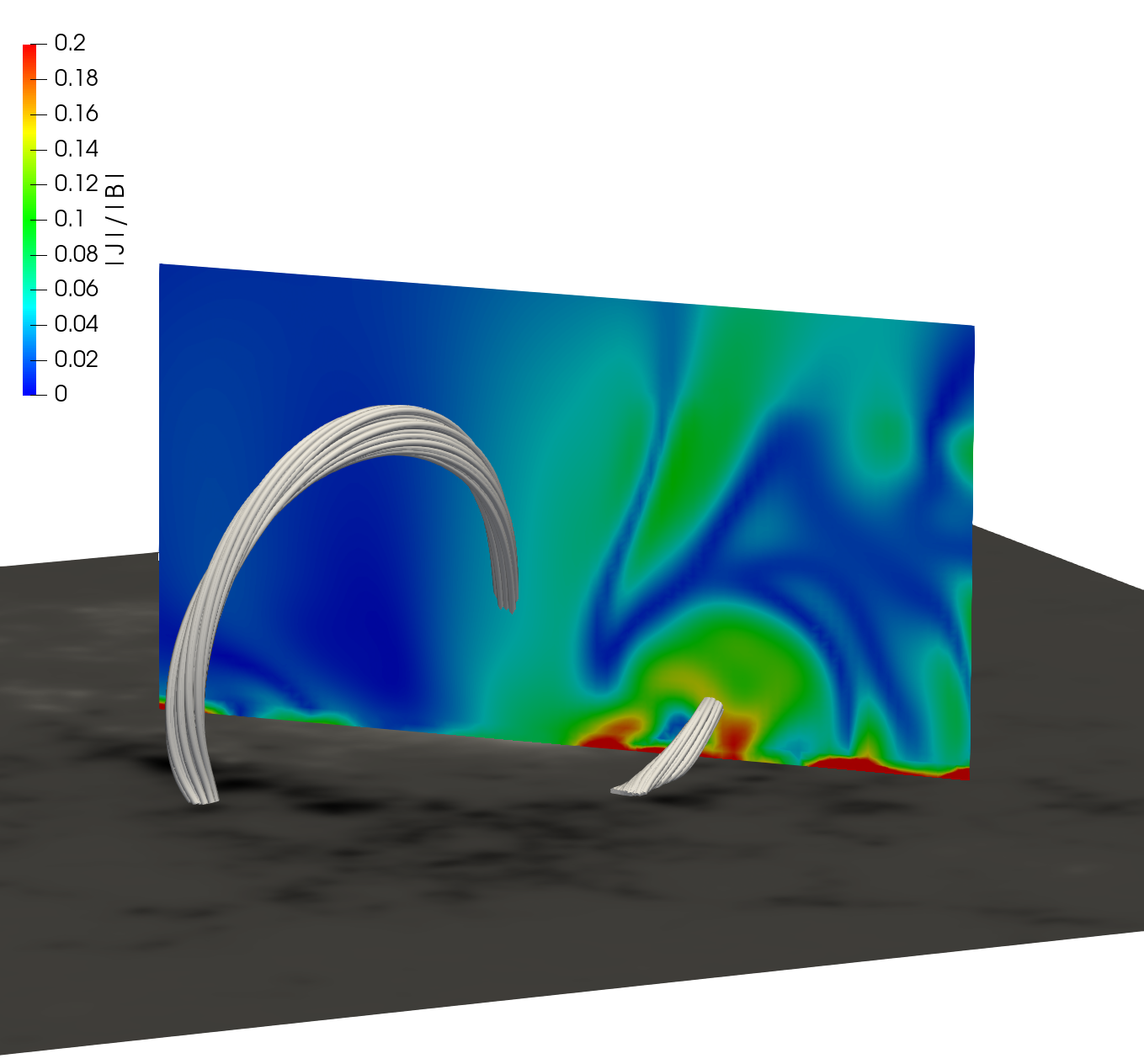}{0.3\textwidth}{(f)}
          }
\caption{Magnetic field topology analysis for event 2: (a-b) selected field lines superimposed on the $B_{z}$ map, and an AIA 131 \r{A} image at 17:00 UT, respectively, the  isosurfaces of (c) $T_{w} = -1$  and (d) $T_{w} = -0.8$  over the background of the AIA 1600 \r{A} observation  at 17:23 UT,  (e) selected field line bundles based on the threshold condition $|T_{w}| > 0.8$ with a composite background of $B_{z}$ map and color-coded flare ribbons (the same as the bottom left panel in Figure \ref{f3}), and (f)  the distribution of $|\mathbf{J}|/|\mathbf{B}|$ on a vertical slice intersecting two field line bundles in (e).} \label{f21}
\end{figure}

For event 2, the configuration of magnetic field lines from the extrapolation result has a good visual correspondence with the AIA observations as shown in Figure \ref{f21}a and \ref{f21}b. There is a clear inverse ``S'' sigmoid structure near the center imaged by AIA 94 \r{A} passband. However, the core field in our extrapolation result mainly consists of several groups of sheared arcades structure over-arched by higher coronal loops, rather than one continuous inverse ``S'' structure (see also \citealt{2017Duan}). Based on the long-term evolution before the eruption, \citet{2015Cheng} found that there was a central sigmoid structure initially appearing in the AIA 94 \r{A} passband at $\sim$15:10 UT and then it had gone through repetitive disappearance and re-appearance processes. So they suggested that a nascent MFR was under formation prior to the major eruption by tether-cutting reconnection. After $\sim$16:55 UT, the sigmoid develops quickly and produces an X1.6 flare and a CME. In order to find a possible MFR structure prior to the flare, we take a look at the $T_{w}$ distribution and find that the majority of the core field region has a negative and relatively small twist number such that $|T_{w}| < 1$. This indicates the absence of a twisted coherent MFR according to the criterion we are using \citep{2016LiuR, 2019Duan}. In Figure \ref{f21}c and \ref{f21}d, we show the isosurfaces of $T_{w} = -1$ and $T_{w} = -0.8$ in the central volume. There is no coherent structure under the $T_{w} = -1$ criterion, though several coherent structures appear for a lower threshold in magnitude $T_{w} = -0.8$. Comparing these field line bundles with the locations of the flare ribbons, there are two coherent weakly twisted field line groups as presented in Figure \ref{f21}e. The left group has a maximum $|T_{w}| \sim 0.82$ but extends to a relatively far away location from the ribbon (also to a height of $z \sim$ 50$''$). It also appears to be nearly perpendicular to the local PIL. Another group of field lines lies close to the bottom boundary and has a maximum $|T_{w}| \sim 0.97$, which still, strictly speaking, fails to satisfy the MFR criterion. In addition, the current density distribution in Figure \ref{f21}f at the intersections of a vertical slice with two field line bundles also shows less clearly-defined concentrations in those field-line regions. So different from the NLFFF extrapolation result by \citet{2016Zhao} and the time-dependent magnetofrictional modeling result by \citet{2021Kilpua}, which demonstrated the existence of a strongly twisted MFR, we do not produce such a pre-existing MFR prior to the flare eruption for event 2. 

\section{Estimation of magnetic properties of MFRs}\label{3}

After the identification of MFRs on the Sun for the two events, we look further into the magnetic properties of MFRs at different stages or locations and try to find a potential correlation among them. For example, the total magnetic flux (generally considered conserved) is one of the most important quantitative properties of MFRs that can be measured or derived to make a connection between CME-MFR and the corresponding ICME/MC \citep{2007Qiu, 2014Hu}. 

Specifically, for the identified pre-existing MFR in event 1, we give below a quantitative description of its magnetic properties. The axial magnetic flux enclosed by the pre-existing MFR's footpoints is calculated for further analysis. Regions of footpoints are obtained by extracting the intersection points between field lines from the MFR and a slice parallel to the bottom boundary (photosphere). We choose a slice at a height of 1.0$''$ where two well-separated groups of footpoints are obtained. In general, more grid points are included under a smaller $|T_{w}|$ threshold value for the MFR criterion, i.e., more points with $|T_{w}|$ exceeding such a value. Here we denote the region dominated by the positive magnetic field in the MFR footpoints as `$FP_{+}$' and the region taken up mainly by the negative magnetic field in the MFR footpoints as `$FP_{-}$'. The total axial (or toroidal) magnetic flux for both regions is calculated based on $\Phi_{z} = \iint B_{z} dS$, where $B_{z}$ is the vertical magnetic field component. The integration is obtained by summing up the magnetic flux from all grid points (pixels) within the identified footpoints regions on the slice. 

Table \ref{footpoints} shows the result of flux calculations of the identified MFR. The differences of $\Phi_{z}$ between $FP_{+}$ and $FP_{-}$ are within one order of magnitude for different $T_{w}$ criteria, though they get smaller for a larger $|T_{w}|$ threshold. Given that the largest $\Phi_{z}$ in magnitude is still significantly smaller than the reconnection flux measured from the flare ribbons after the eruption, especially for the typical criterion $|T_{w}| > 1.0$, we believe that the MFR found from the extrapolation is likely a seed MFR before the eruption. It should be noted that one group of footpoints ($FP_{-}$) locates closer to one main negative polarity of the magnetogram than the other group of footpoints ($FP_{+}$) to any main positive polarity. And $FP_{-}$ takes up a rather smaller area compared to $FP_{+}$, but the latter has much smaller average magnetic field $\langle B_{z}\rangle$, current density $\langle J_{z}\rangle$ and total current $I_{z}$. 

A summary of quantitative results for MFRs in two events is given in Table \ref{summary}. The corresponding in situ modeling results for the two events are provided by \citet{2021Hu} by applying two magnetic cloud reconstruction methods. One of the modeling methods is the Grad-Shafranov reconstruction technique yielding a 2D configuration of the MFR \citep{2002Hu,Hu2017GSreview}. The other method is the optimization approach based on a more general linear force free formulation to obtain a more complex quasi-3D structure \citep{2021Hu,2021Hu2}. For event 1, the twist of the MFR identified in the source region is relatively consistent with the in situ modeling results of the MFR structure, considering the uncertainty of the total twist numbers. The axial magnetic flux calculated from the in situ modeling results is significantly larger than the seed MFR identified in the source region before the eruption, while the reconnection flux measured from the source region after the eruption is generally larger than the axial flux from the in situ modeling results. The poloidal magnetic flux, obtained from the in situ modeling results, appears to agree better with the reconnection flux, subject to the uncertainty in the axial length. Specifically if one assumes an axial length $\in [1, 2]$ au typically (see \citealt{2015Hu}) for an MC flux rope, this amounts to the total poloidal flux in the range 9.2 - 18$\times$10$^{20}$ Mx for event 1, based on the 2D MC modeling result. 

For event 2, the axial (toroidal) flux content from the in situ 3D model agrees with the reconnection flux within their respective uncertainty ranges. Other parameters in the source region are not available (marked by ``...") since we did not find a pre-existing MFR structure before the eruption. The 2D MC model also failed to yield an acceptable solution. The twist of the MFR from the in situ modeling is generally larger than the  twist we found in the groups of  field lines in Figure \ref{f21} (the maximum $|T_{w}| \sim 0.97$). The CME-MFR containing significant amount of flux was likely formed during the eruption through dynamic evolution process in the solar atmosphere. Recently it was  demonstrated by unique observational analysis and data-inspired numerical simulation \citep{2020Xing,2021Jiang2} the ``increase-to-decrease" behavior in the toroidal flux of the CME-MFR. However the applicability of such analysis to our events  is beyond the scope of the current study. To study such a process usually requires discerning multiple flux systems with complex and constantly evolving topologies. And it remains a challenge to separate the toroidal and poloidal flux contents from the reconnection flux, although one pioneering approach developed  by \citet{2009Qiu} for detailed analysis of the reconnection sequence can help in future studies. %\textbf{Such comparison is also consistent with a recent simulation result for the same event by \citet{2021Jiang2} which indicates that the total reconnection flux is much larger than the axial flux in the final MFR given that the reconnected field lines during the later phase result in the increase of the poloidal flux and decrease of the axial (toroidal) flux.}

\section{Conclusions}\label{4}
In this paper, we have identified the MFR structures in the solar source regions and established the connection between MFRs on the Sun and their in situ counterparts quantitatively for two selected events. One event began on 2013 April 11 (event 1, AR 11719) and the other on 2014 September 10 (event 2, AR 12158), respectively. Each event exhibits a sequence of flare, CME and the corresponding ICME observed by multiple space-borne instruments. We perform coronal magnetic field extrapolations for each AR by the CESE-NLFFF-MHD method, which utilizes the preprocessed photospheric magnetograms and the results are also examined through a set of convergence metrics. Remote-sensing observations from SDO are analyzed to find evidence of MFRs and trace the evolution of the associated flares. Specifically we measure the amount of magnetic reconnection flux by analyzing the temporal and spatial evolution of flare ribbons. We combine the extrapolation results with observations to identify MFRs on the Sun before the eruption and estimate their magnetic properties. The main results of our study are summarized as follows.

\begin{enumerate}

\item Observational evidence of MFR footpoints and associated magnetic reconnection flux during the flare eruption are inferred from multi-wavelength observations. From the comparison among EUV observations, there are signs of MFRs for the two events. Based on the flare ribbon measurements, the total magnetic reconnection flux reaches $17 \pm 2.8 \times 10^{20}$ Mx for event 1, and $47\pm 7.5 \times 10^{20}$ Mx for event 2, respectively, which corresponds to the amount of available flux to be injected into the final CME-MFRs.

\item From the combination of extrapolation and observation results, a coherent MFR structure before the flare eruption is identified for event 1. However, there is no pre-existing MFR found for event 2, based on the same set of MFR criteria, including the requirement for the field-line twist number $|T_w|>1.0$ and also both regions of the MFR field-line footpoints close to the main flare ribbons. For event 1, a coherent pre-eruption MFR is determined, which carries a maximum $T_{w} = -1.5$ and its two ends are located near the opposite ends of the respective flare ribbons across the PIL.

\item The magnetic properties of MFRs on the Sun are summarized and compared with their corresponding in situ modeling results from \citet{2021Hu} in Table~\ref{summary}. For event 1, the axial magnetic flux from in situ modeling results is in the order of $10^{20} \sim 10^{21}$ Mx, while the total magnetic reconnection flux after the eruption from the source region is in the order of $\sim 10^{21}$ Mx. Both are significantly larger than the flux in the identified pre-existing MFR's footpoints area which is in the order of $10^{19} \sim 10^{20}$ Mx (see Table~2). 

\item For event 2, there is no pre-existing MFR identified. The amount of the magnetic reconnection flux, $47\pm 7.5 \times 10^{20}$ Mx, agrees with the  corresponding ICME MFR toroidal  flux, $\sim$ 16 - 81 $\times 10^{20}$ Mx, within the limits of the uncertainty ranges. 
\end{enumerate}

These results for the two events indicate the dynamic and complex nature of the MFR formation during its evolution process while some properties (like magnetic flux and twist) are useful for making connections between the formation of MFRs on the Sun and their in situ characteristics in a quantitative manner. Based on these quantitative results, we conclude that the magnetic reconnection process, manifested during solar flares, injects significant amount of magnetic flux into the ensuing CME-MFR. For event 1, the identified pre-existing (or pre-eruption) MFR from the NLFFF extrapolation is likely a seed MFR before the eruption and additional flux is injected through the magnetic reconnection process associated with the flare.
Furthermore, based on the comparison among various inter-related magnetic flux contents and the corresponding flare ribbon morphology for each event, we conclude that for event 1, a quasi-2D configuration of the MFR is largely valid for which the poloidal flux is more meaningfully defined and compared more favorably with the corresponding reconnection flux than the axial flux. For event 2, however, we believe that the MFR topology deviates more from a 2D configuration, but is better described by a quasi-3D model for which the axial flux agrees with the reconnection flux.  In this case, the poloidal flux is not readily defined geometrically because there does not exist a straight field line representing the central axis of a flux rope \citep[see][]{2021Hu}. 
Therefore, for the 3D model, we choose to approximate the poloidal flux by the product of the average field-line twist and the axial flux  \citep[see, e.g.,][]{2014Hu}.

This study represents an effort to make a physical connection between a solar MFR (including the MFR before and after the eruption) and the corresponding ICME/MC by the quantitative comparison of the magnetic properties under different scenarios through extrapolations and observations. It is usually not easy to envisage the existence of a coherent pre-existing MFR, reconstruct it before the eruption in the solar source region for a CME event and make one-to-one connection with its interplanetary counterpart. Efforts have been made continuously on the quantitative description of the MFR configuration with more advanced observations and improved numerical simulation techniques, which is helpful for further understanding the formation and evolution processes of the CME-MFRs. Future studies including more events will be carried out for a deeper understanding of CME-MFRs, where improvements to the extrapolation method and use of high-resolution ground-based data can be implemented. 
%----------------------------------------------------------------------------------

\begin{acknowledgments}
W.H. and Q.H. acknowledge the support by NSF grants AGS-1954503 and AGS-1650854, and the NSO/DKIST Ambassador Program. C.W.J. acknowledges the support from National Natural Science Foundation of China (NSFC 42174200, 41822404), the Fundamental Research Funds for the Central Universities (Grant No. HIT.OCEF.2021033), and Shenzhen Technology Project JCYJ20190806142609035. A.P. would like to acknowledge the support by the Research Council of Norway through its Centres 458 of Excellence scheme, project number 262622, as well as through the Synergy Grant number 810218 459 (ERC-2018-SyG) of the European Research Council. Data for solar observations are provided by GOES, and the HMI and AIA instruments onboard SDO (\url{http://jsoc.stanford.edu/}). The Wind spacecraft measurements are accessed via the NASA CDAWeb: \url{https://cdaweb.gsfc.nasa.gov/index.html/}.

\end{acknowledgments}


% \vspace{5mm}
% \facilities{HST(STIS), Swift(XRT and UVOT), AAVSO, CTIO:1.3m,


% \software{astropy \citep{2013A&A...558A..33A,2018AJ....156..123A},  
%           Cloudy \citep{2013RMxAA..49..137F}, 
%           Source Extractor \citep{1996A&AS..117..393B}
%           }



\appendix
\section*{Convergence Metrics} 
The metrics for the convergence study of the computation during the iteration process are defined below. These include the residual of the field between two successive iteration steps $n$ and $n-1$ ($n > 1$),
\begin{equation}
res^{n}(\textbf{B}) =  \sqrt{\frac{1}{3}\sum_{\delta=x,y,z} \frac{\sum_{i}(B^{n}_{i\delta}-B^{n-1}_{i\delta})^{2}}{\sum_{i}(B^{n}_{i\delta})^{2}}},
\end{equation}
where the subscript $i$ refers to the linear indices of grid points and runs over all grid points in the computational volume. We also estimate the force-freeness of the numerical result by a current-weighted metric, \textit{CWsin}, which is defined by (equivalent to the weighted sine of the angle between \textbf{J} and \textbf{B}),
\begin{equation}
\mbox{\textit{CWsin}} = \frac{\sum_{i} |\textbf{J}_{i}|\sigma_{i}}{\sum_{i} |\textbf{J}_{i}|};
\, \sigma_{i} = \frac{|\textbf{J}_{i} \times \textbf{B}_{i}|}{|\textbf{J}_{i}||\textbf{B}_{i}|}. 
\end{equation}

Other important parameters including the total magnetic energy $E_{tot}$ and the divergence-free condition $\langle|f_{i}|\rangle$ \citep{2000Wheatland, 2008Metcalf} are defined by, 
\begin{equation}
\begin{split}
E_{tot} &= \sum_{i} \textbf{B}_{i}^{2} \Delta V_{i}, \\ \langle|f_{i}|\rangle &= \frac{1}{M}\sum_{i} \frac{(\nabla \cdot \textbf{B})_{i}}{6|\textbf{B}_{i}| / (\Delta x)_{i}},
\end{split}
\end{equation}
where $\Delta x$ is the grid spacing and $M$ refers to the total number of grid points contained, and $\Delta V_{i} = (\Delta x)_{i} \times (\Delta y)_{i} \times (\Delta z)_{i}$ representing a volume element.

% perhaps add another appendix for in situ modeling results of event 2


\bibliography{CME_MFRs}{}
\bibliographystyle{aasjournal}

%%%%%%%%%%%%%%%Tables and Figures%%%%%%%%%%%%%%%%%%%%%%%%%%%%%%%%%%%%%%%%
\begin{deluxetable}{cccccc}
\tablecaption{Timelines of two Flare-CME-ICME events observed by multiple spacecraft (all times are in UT).\label{twoevent}}
\tablewidth{0pt}
\tablehead{
\colhead{Flare peak time} & \colhead{Magnitude} & \colhead{Location} & \colhead{STEREO} & \colhead{SOHO/LASCO} 
&\colhead{WIND/ICME} \\
\nocolhead{Flare peak time} & \colhead{} & \colhead{} & \colhead{CME time} & \colhead{CME time} 
&\colhead{arrival time}
}
\startdata
2013-04-11T07:16&M6.5&N07E13&07:39&07:24&04-14T17:00\\
2014-09-10T17:45&X1.6&N11E05&17:54&18:00&09-12T22:00\\
\enddata
% \tablecomments{sample comment}
\end{deluxetable}

%%%%%%%%%%%%%%%     Event 2013-04-11 %%%%%%%%%%%%%%%%%%%%%%%%%%%%%%%%

%%%%%%%%%%%%%%% Event 2014-09-10 %%%%%%%%%%%%%%%%%%%%%%%%%%%%%%%%%%%%%%%%%%%%%

\begin{deluxetable}{ccccccc}
\tablecaption{Magnetic properties of MFR footpoints in AR11719 for event 1 with different criteria at $z$ = 1.0$''$: $\Phi_{z}$, the sum of normal magnetic flux over all associated grids for each group of identified footpoints; the fourth column shows the number of grids on the chosen slice containing the identified footpoints; $\langle B_{z}\rangle$, the average vertical magnetic field for each group of footpoints; $\langle J_{z}\rangle$, the average current density in the $z$ direction for each group of footpoints, $J_{z} = \frac{(\nabla \times \textbf{B})_{z}}{\mu_{0}}$, and  $I_{z} = \Sigma (J_{z} dS)$, the total current in the $z$ direction for each group. ($+$): positive flux; ($-$): negative flux.}\label{footpoints}
\tablewidth{0pt}
\tablehead{
\colhead{} & \colhead{} & \colhead{$\Phi_{z}$} & \colhead{\# of } & \colhead{$\langle B_{z}\rangle$} 
&\colhead{$\langle J_{z}\rangle $} &\colhead{$I_{z} $} \\
\colhead{} & \colhead{} & \colhead{($10^{20}$ Mx)} &\colhead{Grids}&\colhead{(G)} & \colhead{($10^{-3}$ A$\cdot$m$^{-2}$)} &\colhead{($10^{10}$ A)}
}
\startdata
\multicolumn{7}{c}{\textbf{$|T_{w}| >$ 0.8}}\\
$FP_{+}$&($+$)&0.510&165&58.2&-0.79&-6.9\\
&($-$)&0&&&&\\
% \midrule
$FP_{-}$&($+$)&0&64&-394&5.8&20\\
&($-$)&-1.34&&&&\\
\hline
\multicolumn{7}{c}{\textbf{$|T_{w}| >$ 0.9}}\\
$FP_{+}$&($+$)&0.385&114&63.6&-0.60&-3.6\\
&($-$)&0&&&&\\
% \midrule
$FP_{-}$&($+$)&0&48&-367&5.8&15\\
&($-$)&-0.934&&&&\\
\hline
\multicolumn{7}{c}{\textbf{$|T_{w}| >$ 1.0}}\\
$FP_{+}$&($+$)&0.283&68&78.5&-0.19&-0.68\\
&($-$)&0&&&&\\
% \midrule
$FP_{-}$&($+$)&0&35&-343&5.8&11\\
&($-$)&-0.638&&&&\\
\enddata
\end{deluxetable}

%%%%%%%%%%%%%%%%%%%%%%%%%%%%%%%%%%%%%%%%%%%%%%%%%%%%%%%%%%%%%%%%%%%

\begin{table}
\begin{center}
\caption{Summary of magnetic properties for the MFRs in two events.}\label{summary}
\begin{tabular}{cccc}
\tableline
\tableline
Parameter&Source Region Results&\multicolumn{2}{c}{in situ Modeling Results\tablenotemark{a}}\\
(all flux in $10^{20}$ Mx)&&2D&3D\\
\tableline
\multicolumn{4}{c}{\textbf{Event 1: 2013-04-11}}\\
Axial flux $\Phi_{z}$&0.3 - 1.3 \tablenotemark{b}&5.7&8.9 - 14\\
Twist $\tau$&$\sim$ 1.5\tablenotemark{b} (axis)&1.6 /au&0.84 - 1.1 /au\\
Reconnection flux\tablenotemark{c}&$17 \pm 2.8$&...&...\\
Poloidal flux &...&9.2 /au&10 - 12 /au\\
\hline
\multicolumn{4}{c}{\textbf{Event 2: 2014-09-10}}\\
Axial flux $\Phi_{z}$&...&...&16 - 81\\
Twist $\tau$&... &...&1.0 - 2.4 /au\\
Reconnection flux\tablenotemark{c}&$47\pm 7.5$&...&...\\
Poloidal flux &...&...& 38 - 81 /au\\
\tableline
\tableline
\end{tabular}

\tablenotetext{a}{In situ modeling results of magnetic clouds cited from \citet{2021Hu}. For the 3D model, the poloidal flux is approximated by $\tau\Phi_z$.}
\tablenotetext{b}{Parameters for identified ``pre-existing" MFR only.}
\tablenotetext{c}{The reconnection flux is estimated based on Figures \ref{f2} and \ref{f3}.}
\end{center}
\end{table}



\end{document}